\documentclass[twoside,11pt]{article}
\usepackage{nonatbib}
\usepackagewithoutnatbib{jmlr2e}
\usepackage{fixthmtools}
\usepackage[T1]{fontenc}
\usepackage{microtype}
\usepackage{hyperref}
\usepackage{lastpage}
\usepackage{subcaption}

\usepackage[authoryear,citations,centerfigures,commands,enumerate,environments,lmrscinnershape,theorems]{AVT}

\bibliography{references}
\let\UrlFont\scshape

\usepackage{tikz}

\renewcommand{\Re}{\operatorname{Re}}

\DeclareMathOperator{\GL}{GL}
\DeclareMathOperator{\Ort}{O}
\DeclareMathOperator{\SL}{SL}
\DeclareMathOperator{\SO}{SO}
\DeclareMathOperator{\SU}{SU}

\DeclareMathOperator{\SPD}{SPD}
\DeclareMathOperator{\SSPD}{SSPD}

\declaretheorem[style=plain]{Assumption}

\declaretheoremstyle[headformat=\NAME]{customstyle}
\declaretheorem[style=customstyle,name={\Cref*{thm:spd_sampling}-\ensuremath{\v\eps}},Refname={\Cref*{thm:spd_sampling}-\ensuremath{\eps}}]{CustomLemma}
\Crefname{equation}{}{}

\title{Stationary Kernels and Gaussian Processes on Lie Groups and their Homogeneous Spaces II: non-compact symmetric spaces}

\author{\name Iskander Azangulov\textsuperscript{\ensuremath{*}}
\email iska.azn@gmail.com \\
\addr St. Petersburg State University and University of Oxford
\AND
\name Andrei Smolensky\textsuperscript{\ensuremath{*}}
\email andrei.smolensky@gmail.com \\
\addr St. Petersburg State University and Neapolis University Pafos
\AND
\name Alexander Terenin
\email avt28@cornell.edu \\
\addr University of Cambridge and Cornell University
\AND
\name Viacheslav Borovitskiy
\email viacheslav.borovitskiy@gmail.com \\
\addr ETH Zürich
}

\jmlrheading{25}{2024}{1-\pageref{LastPage}}{1/23; Revised
7/24}{8/24}{23-0115}{Iskander Azangulov, Andrei Smolensky, Alexander Terenin, Viacheslav Borovitskiy}
\ShortHeadings{\textls*[-2]{Stationary Kernels and Gaussian Processes on Lie Groups and Homogeneous Spaces II}}{Azangulov, Smolensky, Terenin, Borovitskiy}

\editor{Philipp Hennig}

\begin{document}

\maketitle

\begin{table}[b!]
\vspace*{-1.5ex}
\footnoterule
\footnotesize\textsuperscript{\ensuremath{*}}Joint first author.
\hfill\null
\end{table}

\begin{abstract}
Gaussian processes are arguably the most important class of spatiotemporal models within machine learning.
They encode prior information about the modeled function and can be used for exact or approximate Bayesian learning.
In many applications, particularly in physical sciences and engineering, but also in areas such as geostatistics and neuroscience, invariance to symmetries is one of the most fundamental forms of prior information one can consider.
The invariance of a Gaussian process' covariance to such symmetries gives rise to the most natural generalization of the concept of stationarity to such spaces.
In this work, we develop constructive and practical techniques for building stationary Gaussian processes on a very large class of non-Euclidean spaces arising in the context of symmetries.
Our techniques make it possible to (i) calculate covariance kernels and (ii) sample from prior and posterior Gaussian processes defined on such spaces, both in a practical manner.
This work is split into two parts, each involving different technical considerations: part I studies compact spaces, while part II studies non-compact spaces possessing certain structure.
Our contributions make the non-Euclidean Gaussian process models we study compatible with well-understood computational techniques available in standard Gaussian process software packages, thereby making them accessible to practitioners.
\end{abstract}


\section{Introduction} \label{sec:intro}

This work is the second part of a two-part series of papers that aims at bringing Gaussian process regression into the machine learning toolbox for a large class of Riemannian manifolds admitting analytical descriptions.
To do so, in part I \cite{part1} we studied a rich class of compact manifolds, including compact \emph{Lie groups} such as the special orthogonal group $\SO(n)$ and the special unitary group $\SU(n)$, as well as their \emph{homogeneous spaces}, such as the sphere $\bb{S}_n$ and Stiefel manifold $\f{V}(k, n)$.
We developed computational techniques based on suitable generalizations of Bochner's Theorem due to \textcite{yaglom1961}, which express stationary kernels on a compact homogeneous space $X = G/H$ as series in terms of \emph{characters} of irreducible unitary representations of the group $G$, or the respective \emph{zonal spherical functions}.
These expressions generalize a number of previous works devoted to extending Gaussian processes to non-Euclidean spaces \cite{lindgren11,coveney20,borovitskiy2020,borovitskiy2021,borovitskiy2022,hutchinson2021,dacosta2023}, and give a comprehensive mathematical treatment on how to implement stationary kernels and Gaussian processes on compact homogeneous spaces. 
These spaces occur in many application areas ranging from neuroscience~\cite{jensen2020} to dynamical systems and robotics~\cite{jaquier2020,jaquier2022}, as well as many others, and provide a rich setting for studying how geometry affects machine learning algorithms such as kernel methods \cite{rosa23}.
For further motivation and a list of applications and examples, please see the introduction to part I.

Many application areas, such as for instance robotics and neuroscience, give rise to \emph{non-compact} manifolds such as hyperbolic spaces, and spaces of positive definite matrices \cite{jaquier2022, jaquier2024, nejatbakhsh2024}.
For these, comprehensive computational techniques suitable for performing Gaussian process regression, and downstream techniques that rely on it such as Bayesian optimization, have yet to be developed. 
Our goal in this work is to develop them.

One of the key challenges when $G$ is non-compact is that one must work with geometric analogs of the Fourier transform, rather than of Fourier series.
Moreover, the setting is delicate: one can show that (i) stationary kernels on general non-compact non-abelian Lie groups usually do not exist, in an appropriate sense, and (ii) non-compact abelian Lie groups, provided they are connected, are isomorphic to products of Euclidean spaces with tori, which can already be handled by the combination of classical Euclidean techniques and compactness-based ideas developed in part I.
Then, a product of kernels can be used to define a reasonable kernel on the corresponding product space, obtaining a reasonable Gaussian process model.
This motivates us to work with a subset of homogeneous spaces, namely \emph{symmetric spaces}, which are in some sense the right class of spaces on which one can reasonably expect Bochner-like results to hold in a concrete-enough manner to be amenable to numerical~methods.

As in part I, we will begin with an abstract description of stationary kernels, and develop numerical techniques to make these kernels into concrete and practically viable machine learning models.
Specifically, we develop the Bochner-type results of \textcite{yaglom1961} into concrete numerical methods, using the \emph{spherical Fourier transform} of \textcite{helgason1962differential} for symmetric spaces.
Our developments reveal that such kernels admit a maximum correlation value, meaning that their respective Gaussian processes are non-constant in the large length-scale limit---in sharp contrast with what happens in the Euclidean and compact settings.

For the first time, our work enables one to efficiently implement both the squared exponential (also known as heat, diffusion, RBF) and Matérn kernel classes in, for instance, hyperbolic space and the space of positive definite matrices.
We study these cases in additional detail, and propose specific techniques that leverage their individual properties to enable more efficient approximations, compared to the general case.
These developments reveal unanticipated connections between geometric kernels and random matrix theory.
As part of our work, we contribute implementations of these ideas to the \href{https://geometric-kernels.github.io}{\textsc{GeometricKernels}}\footnote{Available at \url{https://geometric-kernels.github.io}. A prototypical software implementation, as well as our experiments, can be found at \url{https://github.com/imbirik/LieStationaryKernels}.}~library.

\paragraph{Notation}
\label{sec:intro:goals_structure_notation}
Following part I, we use bold italics ($\v{a}$, $\v{b}$) to simultaneously refer to finite-dimensional vectors in $\R^n$, and to elements of the product space $X^n$ for a given set $X$.
We use bold upface letters ($\m{A},\m{B}$) to refer to matrices.
If $f: X \-> Y$ is a function, then for $\v{x} \in X^n$ we write $f(\v{x})$ to refer to the vector $\del{f(x_1), \ldots, f(x_n)}^\top$.
For a two-argument function $k: A \x B \-> C$, and $\v{a} \in A^n$, $\v{b} \in B^m$, the notation $\m{K}_{\v{a} \v{b}}$ refers to the $n \x m$ matrix with entries $k(a_\ell, b_{\ell'})$ for $1 \leq \ell \leq n$ and $1 \leq \ell' \leq m$.
We refer to conjugation of complex numbers by $\conj{x}$, and to the conjugate transpose of a matrix by $\m{A}^* = \conj{\m{A}^\top}$.

\subsection{Gaussian Processes and Gaussian Process Regression}
\label{sec:intro:gpr}

A \emph{Gaussian process} $f$ is a random function mapping some set $X$ to the reals, whose finite dimensional distributions are all jointly Gaussian.
One can show that every Gaussian process is uniquely determined by its mean function $m$ and covariance kernel $k$ given by
\[
m(x) &= \E f(x)
&
k(x, x') &= \Cov \del{f(x), f(x')}.
\]

Gaussian processes are among the most commonly used priors for unknown functions in Bayesian learning.
If $f \~[GP](0, k)$ and we observe the dataset $\v{x},\v{y}$, where $y_i = f(x_i) + \eps_i$ with $\v\eps \~[N](\v{0}, \m\Sigma_{\eps})$, then the posterior $f \given \v{y}$ is also a Gaussian process admitting the explicit form \cite{chiles2009,wilson2020,wilson2021}
\[ \label{eqn:pathwise_cond}
(f \given \v{y})(\.) &= f(\.) - \m{K}_{(\.)\v{x}} (\m{K}_{\v{x}\v{x}} + \m\Sigma_{\eps})^{-1} (\v{y} - f(\v{x}) - \v\eps)
&
\v\eps &\~[N](\v{0},\m\Sigma_{\eps})
\]
to which we refer to as \emph{pathwise conditioning}.
From this expression, one can compute the posterior's mean and covariance, obtaining predictions and their associated uncertainty.

By using \Cref{eqn:pathwise_cond}, one can obtain samples from the posterior by generating samples from the prior and substituting them into~\Cref{eqn:pathwise_cond} as $f(\.)$ and $f(\v{x})$.
Using this, if one is able to approximately sample from the prior $f$ with a given computational complexity with respect to the number of evaluation locations, it follows one can approximately sample from the posterior with the same evaluation complexity.

\subsection{Gaussian Processes That are Stationary Under Group Action}
\label{sec:intro:stationary}

Suppose there is a group\footnote{See \textcite{kondor2008} or \textcite{robinson2008} for a mild or more technical introduction to groups, respectively.} $(G,\bdot)$ acting on $X$, with group action $\lacts : G \x X \-> X$.
A Gaussian process $f \~[GP](0, k)$ on $X$ is called \emph{stationary} if its distribution is invariant under group action.
This means that for every finite set of points $x_1, \ldots, x_m \in X$ and all $g \in G$, the random vectors $\del{f(x_1), \ldots, f(x_m)}$ and $\del{f(g \lacts x_1), \ldots, f(g \lacts x_m)}$ have the same distribution.
It is easy to see that this is equivalent to 
\[
k(g \lacts x, g \lacts x') = k(x, x').
\]
When a kernel $k$ satisfies the property above, it is termed \emph{stationary} with respect to the group $G$ and action $\lacts$.

Stationarity under group action includes the classical case $X = \R^n$, where $G = \R^n$ and the group action is $g \lacts x = g + x$, but is much richer as it allows one to consider general spaces and groups, far beyond Euclidean shift-invariance.
For example, one obtains isotropic Gaussian processes by replacing the translation group $G = \R^n$ with the Euclidean group $G = \f{E}(n)$ acting on $\R^n$ with rotations and reflections in addition to translations.

For stationary processes under the action of sufficiently regular and rich group of symmetries, one can show Bochner-type theorems. 
In the Euclidean case, this is
\[
k(\v{x}, \v{x}') = \int_{\R^n} e^{2\pi i \v{\lambda}^\top(\v{x}-\v{x}')}  \d S(\v{\lambda}) = \int_{\R^n} a^{(\v{\lambda})} e^{2\pi i \v{\lambda}^\top(\v{x}-\v{x}')}  \d\v{\lambda}
\]
where, assuming $S$ is absolutely continuous, $a^{(\v{\lambda})}$ is the density of $S$, and $\int_{\R^n} a^{(\v{\lambda})} \d \v{\lambda} < \infty$.

Another form of Bochner's theorem is available in the case of a compact \emph{homogeneous space} $X$.
Deferring a formal definition, we will for now say that this is a compact smooth manifold with a sufficiently-rich group $G$ of symmetries acting on it.
If we fix an arbitrary point $x_0 \in X$, which is analogous to a sphere's pole, and define $g_1, g_2 \in G$ in a way that ensures $g_1 \lacts x_0 = x$ and $g_2 \lacts x_0 = x'$---where such $g_1, g_2$ exist because $G$ is sufficiently rich---the appropriate analog of Bochner's theorem~\cite{part1, yaglom1961} says
\[ \label{eqn:bochner_comp}
k(x, x')
=
\sum_{\lambda\in\Lambda} \sum_{j,k=1}^{r_\lambda} a^{(\lambda)}_{jk} 
\Re \pi^{(\lambda)}_{jk} (g_2^{-1} \bdot g_1)
\]
where the matrices $\m{A}^{(\lambda)}$ with entries $a^{(\lambda)}_{jk}$ must be positive semi-definite and satisfy the condition ${\sum_{\lambda \in \Lambda} \tr{\m{A}^{(\lambda)}} < \infty}$.
The two key changes compared to the Euclidean case are: (i) instead of an integral, we have a sum over the countable set $\Lambda$ of irreducible unitary representations of the group $G$, where the corresponding weights $\m{A}^{(\lambda)}$ can be understood as a kind of spectral measure, which is now \emph{discrete} and \emph{matrix-valued}, and (ii) the complex exponentials are replaced with the \emph{zonal spherical functions} $\pi^{(\lambda)}_{jk}: G \-> \C$.
The zonal spherical functions $\pi^{(\lambda)}_{jk}$ and the values $r_{\lambda}$ are determined by the space $X$ itself, while $\Lambda$ is determined solely by the group $G$.

The expression~\Cref{eqn:bochner_comp} simplifies greatly when $r_{\lambda} \leq 1$, and the inner sum either collapses to one term, or vanishes.
In this case, the matrices $\m{A}^{(\lambda)}$ simplify to scalars $a^{(\lambda)}$, which now behave much like a discrete analog of the spectral measures in the Euclidean case.
This simplification occurs for spaces termed \emph{symmetric spaces} \cite{yaglom1961}---these include compact Lie groups acting on themselves from both sides \cite{terras2016}, as well as homogeneous spaces of constant sectional curvature, such as for instance spheres or real projective spaces \cite{yaglom1961}, or hyperbolic spaces in the non-compact setting.

In the non-compact case, general homogeneous spaces are much less well-behaved than in the compact case.
Here, even in the seemingly-simplest case of $X$ being a regular non-compact non-abelian Lie group, non-trivial stationary Gaussian processes do not exist.

\begin{result}
Let $G$ be a non-compact simple\footnote{\emph{Simple Lie groups} are the basic building blocks for general connected Lie groups. See the formal definition in~\Cref{apdx:noncompact}.} non-abelian Lie group.
Then the only kernels on $G$ stationary with respect to both left and right action of $G$ on itself are identically constant.
\end{result}

\begin{proof}[Proof sketch]
Here, we sketch the argument, without introducing the rather heavy abstract machinery needed to make the statement mathematically rigorous: the omitted technicalities may be recovered from \textcite[pp. 601--602]{yaglom1961}.
Recall that every Lie group can be viewed as a homogeneous space, where $H = \{e\}$ is the trivial group.
Under this viewpoint, one can show that every left-stationary kernel must satisfy an analog of \Cref{eqn:bochner_comp}, but where sums are replaced with abstract integrals, and $\m{A}^{(\lambda)}$ is replaced with a suitable infinite-dimensional generalization.
A mirrored expression holds for right-stationary kernels, which forces us to choose a form akin to $\m{A}^{(\lambda)} = a^{(\lambda)} \m{I}$, but this results in a statement similar to
\[ \label{eqn:tr_sum_proof}
\sum_{\lambda \in \Lambda}
\tr{\m{A}^{(\lambda)}}
=
\sum_{\lambda \in \Lambda} 
a^{(\lambda)}
r_{\lambda}
<
\infty.
\]
When a homogeneous space is a group, each $r_{\lambda}$ is equal to the dimension $d_{\lambda}$ of the unitary representation corresponding to $\lambda$.
Crucially, for non-compact simple non-abelian Lie groups, all unitary representations except the trivial one are infinite-dimensional \cite{segalvonneumann1950}.
This implies $r_{\lambda} = d_{\lambda} = \infty$ for all $\lambda$ corresponding to non-trivial representations, making it impossible for \Cref{eqn:tr_sum_proof} to hold unless $a^{(\lambda)} = 0$ holds for them.
Since the only remaining representation is the trivial representation, setting a non-zero $a^{(\lambda)}$ for this representation gives the class of constant kernels.
\end{proof}

If we consider symmetric spaces, then $r_{\lambda} \leq 1$ holds even in the non-compact setting.
This enables a clean generalization of Bochner's theorem. 
Moreover, many interesting non-compact homogeneous spaces, such as hyperbolic space and certain spaces of matrices, are symmetric.
This makes one wonder: are non-compact symmetric spaces a good setting in which to develop numerical techniques needed to use Gaussian-process-based models?
We proceed to answer this question affirmatively in the sequel.

\section{Stationary Gaussian Processes on Symmetric Spaces} \label{sec:stationary}

To begin, we recall the formal definition of \emph{homogeneous spaces}, which can be formulated in either geometric or algebraic terms.
Geometrically, a homogeneous space is a smooth manifold\footnote{Such a manifold often possesses a canonical (pseudo-)Riemannian structure, which we will discuss later.} $X$, equipped with a transitively acting Lie group $G$ consisting of diffeomorphisms of $X$.
Here, \emph{transitive} means that for all $x, x' \in X$ there is a $g \in G$ with $g \lacts x = x'$.
We call $G$ the group of symmetries of $X$.
$X$ can be represented as the quotient $X = G / H$ where $H = \cbr{g \in G : g \lacts x_0 = x_0}$ for an arbitrary $x_0 \in X$ is termed the \emph{isotropy group} \cite[Proposition 4.2]{arvanitogeorgos2003}. 
This implies that elements of $X$ can be regarded as \emph{cosets} $g \bdot H$, $g \in G$: we can express $x \in G / H$ as $g \bdot H$ where $g \in G$ satisfies $g \lacts x_0 = x$.
This is the algebraic definition of a homogeneous space: a quotient of a Lie group $G$ by any of its subgroups $H \subseteq G$.
There is a canonical smooth manifold structure on any such quotient \cite[Proposition 4.1]{arvanitogeorgos2003} and one can show that $G$ acts transitively on~$G/H$.

A manifold $X$ can potentially be given the structure of a homogeneous space in more than one way, by choosing different groups $G$.
For example, consider the Euclidean group $\f{E}(n)$, acting transitively on $\R^n$ by translations, rotations, and reflections.
If we fix $\v{0} \in \R^n$ as $x_0$, its isotropy group will be exactly the group $\Ort(n)$ generated by rotations and reflections, thus $\R^n = \f{E}(n) / \Ort(n)$.
On the other hand, the subgroup of $\f{E}(n)$ that consists of translations alone already acts transitively on $\R^n$, and is isomorphic to the addition group of $\R^n$ itself, meaning that $\R^n = \R^n / \cbr{\v{0}}$ where $\cbr{\v{0}}$ is the trivial subgroup of $\R^n$ consisting of only the identity element.

We now define \emph{symmetric spaces}.
Geometrically, a symmetric space is a Riemannian manifold whose isometry group contains an inversion symmetry about every point.
This means that for every $x \in X$ there is an isometry $g_x: X \-> X$ which inverts the direction of all of the geodesics passing through~$x$.
From this definition, one can show that a symmetric space is necessarily also a homogeneous space \cite[Theorem 6.2]{arvanitogeorgos2003}.
A symmetric space can also be defined algebraically as a homogeneous space $G/H$ where $G$ is a connected Lie group for which there exists an automorphism $s : G \-> G$ which is an involution, namely $s(s(g)) = g$ for all $g \in G$, and $H$ is the union of connected components of the subgroup $G^{s} = \cbr{g \in G : s(g) = g}$.
These definitions are equivalent, provided the algebraically-defined symmetric space is equipped with a Riemannian metric which is left-$G$-invariant~\cite[Theorem 6.3]{arvanitogeorgos2003}.
In this work, we always assume conditions which imply $H = G^s$, or equivalently that $s(g) = g$ if and only if $g \in H$.\footnote{Note that \textcite{yaglom1961} also works with this slightly more restrictive notion of a symmetric space.}

We now consider the analog of Bochner's Theorem for non-compact symmetric spaces.
As noted before, if $G$ is non-compact, one must work with general analogs of the Fourier transform instead of Fourier series.
At best, one can therefore hope to represent stationary kernels through integrals over the index set $\Lambda$ of irreducible unitary representations, which is uncountable.
With these notions, one can obtain an integral expression for symmetric spaces which is similar to the Euclidean case, as long as $H = G^s$.

\begin{theorem} 
\label{thm:stationary_symmetric}
Let $X = G/H$ be a non-compact symmetric space, where $G$ is a Lie group of type I.\footnote{We do not dwell on the technical definition of a \emph{type I} group, which involves $C^*$-algebraic considerations and implies that all irreducible unitary representations are uniquely characterized by characters, which is usually the case: our subsequent assumptions will imply $G$ being type I. See \textcite[p. 599]{yaglom1961} and \textcite{naimark1959} for details.}
A Gaussian process $f \~[GP](0, k)$ is stationary on $X$ if and only if $k$ is of form
\[ \label{eqn:stationary_symmetric:kernel}
k(x, x') = k(g_1 \bdot H, g_2 \bdot H) = \Bbbk(g_2^{-1} \bdot g_1) = \int_{\Lambda} \pi^{(\lambda)}(g_2^{-1} \bdot g_1) \d\mu_k(\lambda)
\]
for a function $\Bbbk : G \-> \R$, where $g_1 \bdot H$ and $g_2 \bdot H$ are the cosets of $x$ and $x'$, $\Lambda$ is the set of all---not necessarily finite dimensional---irreducible unitary representations of $G$ with $r_{\lambda} = 1$, $\pi^{(\lambda)}$ are the zonal spherical functions, and $\mu_k$ is a nonnegative finite measure over~$\Lambda$.
\end{theorem}

\begin{proof}
\textcite[Theorem 6]{yaglom1961}.
\end{proof}

\begin{figure}[t!]
\begin{subfigure}{0.24\textwidth}
\begin{tikzpicture}
\path[use as bounding box] (-1.75,-1.25) rectangle (1.875,3.75);
\node at (0,2.5) {\includegraphics[scale=0.25]{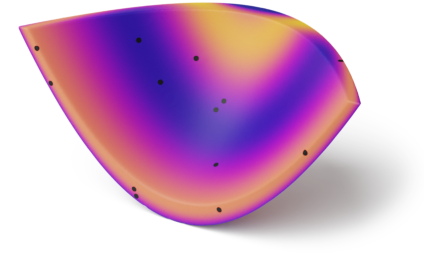}};
\node at (0,0) {\includegraphics[scale=0.25]{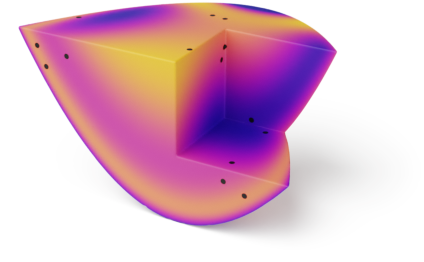}};
\end{tikzpicture}
\caption{Ground truth}
\end{subfigure}
\begin{subfigure}{0.24\textwidth}
\begin{tikzpicture}
\path[use as bounding box] (-1.75,-1.25) rectangle (1.875,3.75);
\node at (0,2.5) {\includegraphics[scale=0.25]{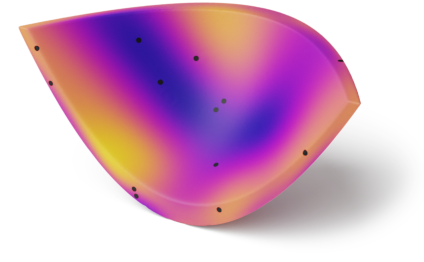}};
\node at (0,0) {\includegraphics[scale=0.25]{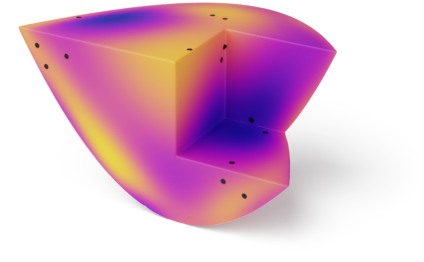}};
\end{tikzpicture}
\caption{Mean}
\end{subfigure}
\begin{subfigure}{0.24\textwidth}
\begin{tikzpicture}
\path[use as bounding box] (-1.75,-1.25) rectangle (1.875,3.75);
\node at (0,2.5) {\includegraphics[scale=0.25]{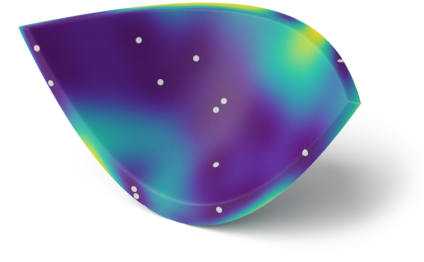}};
\node at (0,0) {\includegraphics[scale=0.25]{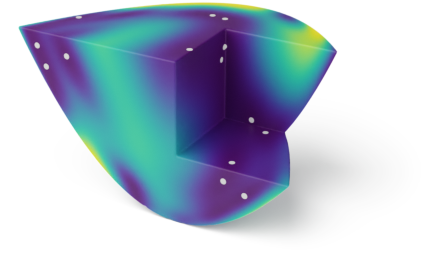}};
\end{tikzpicture}
\caption{Standard dev.}
\end{subfigure}
\begin{subfigure}{0.24\textwidth}
\begin{tikzpicture}
\path[use as bounding box] (-1.75,-1.25) rectangle (1.875,3.75);
\node at (0,2.5) {\includegraphics[scale=0.25]{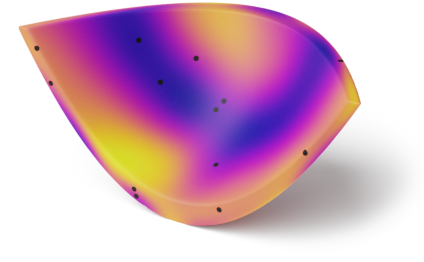}};
\node at (0,0) {\includegraphics[scale=0.25]{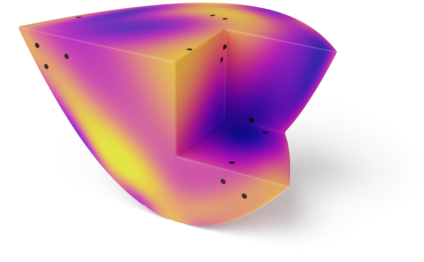}};
\end{tikzpicture}
\caption{Posterior sample}
\end{subfigure}
\caption{Posterior on the manifold $\SPD(d)$ of symmetric positive definite $d \x d$ matrices, equipped with the affine-invariant metric. Here, $d=2$ and $\SPD(d)$ is represented as a cone: see \Cref{sec:spd}. The top and the bottom rows correspond to two cross-sections thereof.}
\label{fig:spd-posterior}
\end{figure}
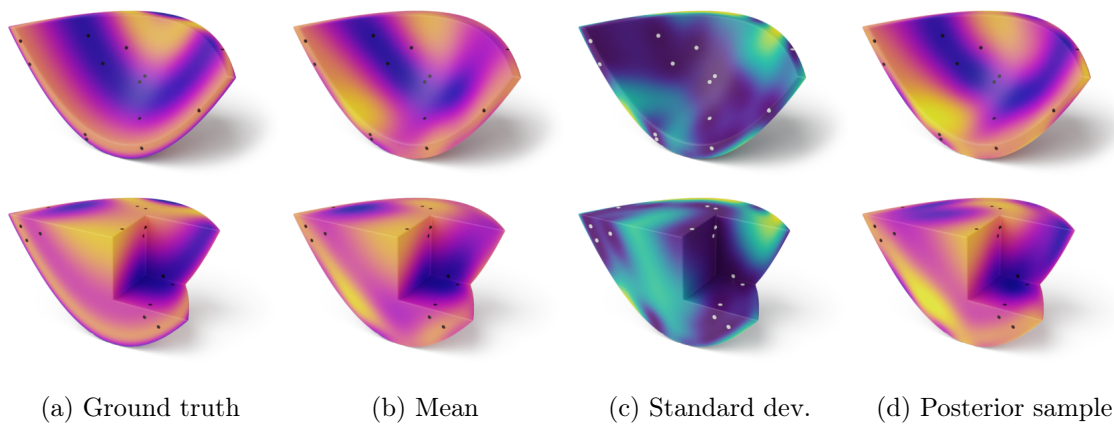

Subsequently, when referring to irreducible unitary representations, we implicitly refer only to those for which $r_\lambda = 1$---as opposed to those for which $r_\lambda = 0$.
An illustration of a posterior Gaussian process, computed using approximations built from this result and presented in the sequel, is given in \Cref{fig:spd-posterior}.
We proceed to study how to concretely calculate expressions of this type for various symmetric spaces of interest.

\section{Computational Techniques for Stationary Gaussian Processes} \label{sec:computation}

We now develop computational techniques for working with the stationary Gaussian processes described in the preceding section.
This involves the following two key computational primitives:
\1*[(a)] Evaluate the covariance kernel $k(\.,\.)$ pointwise.
\2*[(b)] Efficiently sample from the prior $f(\.)$.
\0*
Note that both (a) and (b) generally need to be performed in an automatically-differentiable manner, for instance to optimize model hyperparameters using maximum marginal likelihood.
For a given set of evaluation locations, $f$ can be sampled by forming and factorizing the kernel matrix at cubic cost with respect to the number of evaluation locations: we use the term \emph{efficient sampling} to refer to sampling $f$ approximately \emph{without these costs}.

We focus on these primitives because Gaussian process priors are often used as building blocks for defining larger models.
Here, most numerical methods require either (a) or (b) as part of their steps.
For example, in Hamiltonian Monte Carlo, (a) is needed to compute the posterior density and evaluate the gradient of its logarithm.
Similarly, in variational inference, (b) can be needed to compute stochastic estimators of expectations which appear in the variational objective.
Computing Monte-Carlo-type acquisition functions \cite{wilson18,neiswanger21} in Bayesian optimization, active learning, or other automated decision making applications further requires one to automatically differentiate both expressions---for instance, to perform Thompson sampling in an efficient and numerically stable manner using pathwise conditioning \cite{wilson2020,wilson2021,terenin2024}.
These primitives therefore enable one to apply Gaussian processes as part of a wide class of learning and decision-making algorithms.

The general form of a stationary kernel on a symmetric space $G/H$ under a non-compact group $G$ is given by~\Cref{eqn:stationary_symmetric:kernel}.
From this, it is clear that to approximately compute such a kernel pointwise it is enough to be able to sample from the measure $\mu_k$ and compute the zonal spherical functions~$\pi^{(\lambda)}$.
Letting $\sigma^2 = \mu_k(\Lambda)$, we obtain the random Fourier feature expansion
\[ \label{eqn:simple_noncompact_approx}
&
k(g_1 \bdot H, g_2 \bdot H)
\approx
\frac{\sigma^2}{L}
\sum_{l=1}^L
\pi^{(\lambda_l)}(g_2^{-1} \bdot g_1)
&
\lambda_l &\~
\frac{1}{\sigma^2}\mu_k
.
\]
To transform this into a bona-fide computational approach, we need to describe how to sample from $\mu_k$, which is supported on the set of irreducible unitary representations $\Lambda$, and how to compute $\pi^{(\lambda)}$.
Furthermore, since this expression does not directly factorize into an inner product, for efficient sampling and to guarantee positive semi-definiteness we will need to go beyond \Cref{eqn:simple_noncompact_approx}, leveraging specific algebraic properties of the zonal spherical functions~$\pi^{(\lambda_l)}$.

The high-level idea will be to reduce the general case to the simplest possible non-compact setting, namely Euclidean space---mirroring how, in the compact case, the high-level idea was to reduce the general case to the simplest possible compact setting, namely the torus \cite[Section 3.1]{part1}.
In this section, we assume henceforth the kernel and corresponding spectral measure $\mu_k$ are fixed: we study how to calculate spectral measures for specific kernel classes such as the Matérn and heat kernel classes in \Cref{sec:heat_matern}.

In \Cref{sec:computation:prelim}, we formalize the intuition of the preceding paragraph.
Then, in \Cref{sec:computation:pointwise}, we present a way to bijectively map a certain subset of $\Lambda$---which is sufficiently large to describe all square-integrable kernels---into a Euclidean space, and give an expression for the zonal spherical functions $\pi^{(\lambda)}$---in total, this allows us to evaluate the kernel pointwise.
Following this, in \Cref{sec:computation:sampling} we show how to leverage a certain very special property of the zonal spherical functions to obtain an efficient sampling strategy.
As a byproduct, this allows us to build an alternative approximation to the kernel which, unlike the approximation we introduce in \Cref{sec:computation:pointwise}, is guaranteed to be positive semi-definite---an important property to ensure that downstream algorithms work well in practice.

Before starting this, we note that quotients $G/H$ of a sufficiently regular Lie group $G$ by a sufficiently nice subgroup $H$, including the symmetric spaces we study, admit a \emph{canonical} Riemannian structure---something that will be essential for~\Cref{sec:heat_matern}.
To ensure this and to satisfy assumptions of the spherical Fourier transform theory of Helgason and Harish-Chandra---the theory that will enable us to make the abstract expressions we have seen so far into something concrete and amenable to computation---we assume the following technical conditions.

\begin{Assumption}
\label{asm:symmetric_space_assumptions}
In the symmetric space $G/H$, $G$ is a \emph{semi-simple}\footnote{Note that, if $G$ is semi-simple, it is automatically of type I, see \textcite{harishchandra1953, yaglom1961}.} Lie group with \emph{finite center}, and $H$ is a \emph{maximal compact subgroup} of $G$.
\end{Assumption}

See \Cref{apdx:noncompact} for definitions.
In particular, hyperbolic spaces and the spaces of symmetric positive definite matrices with unit determinant arise in this way and satisfy the regularity assumptions.
One can relax this assumption to situations where the symmetric space of interest is a Riemannian product of a compact space, Euclidean space, and symmetric space satisfying \Cref{asm:symmetric_space_assumptions}: this factorization can be viewed as a crude and less-general analog of the basic classification results for symmetric spaces, see for instance \textcite[Chapter V]{helgason1962differential}. 
We will use this factorization idea in our examples to handle the space of symmetric positive definite matrices in \Cref{sec:spd}, which does not satisfy \Cref{asm:symmetric_space_assumptions}, but to ease presentation work with \Cref{asm:symmetric_space_assumptions} otherwise.

\subsection{Enumerating Representations of Symmetric Spaces}
\label{sec:computation:prelim}

In the preceding section, we formulated a strategy for defining random Fourier features on symmetric spaces, where the high-level idea was to reduce the general case to the simplest possible non-compact setting---that is, Euclidean space, which is abelian.
To actually carry this out, we will need a certain decomposition of $G$.

This is the \emph{Iwasawa decomposition}, which can be viewed as a basis-free generalization of the well-known QR decomposition of a matrix.
If $G = \GL(\R^n)$---we note that in this special case the result still holds, despite $G$ not being semi-simple---then the Iwasawa decomposition can be reinterpreted as saying that the matrix corresponding to an arbitrary group element $g \in G$ can be represented as the product of a unit lower triangular matrix $N(g)$, positive diagonal matrix $A(g)$, and orthogonal matrix $H(g)$.
The Iwasawa decomposition therefore generalizes the QR decomposition, up to transposition, with $Q^{\top}$ corresponding to the product $N(g) A(g)$.
A formal discussion of the Iwasawa decomposition can be found in~\Cref{apdx:noncompact}.

The key reason for introducing this decomposition in our setting is that it turns out, in order to compute a stationary kernel, it suffices---roughly speaking---to integrate only over the abelian, resp. diagonal, part of $G$ in Iwasawa decomposition.
For a Lie group $G$, define the \emph{Lie algebra} $\fr{g}$ to be the tangent space at the identity element equipped with appropriate algebraic structure---see \Cref{apdx:noncompact} for details.
We now state the result.

\begin{proposition}
\label{thm:iwasawa_support}
Let $G/H$ be a symmetric space satisfying \Cref{asm:symmetric_space_assumptions}.
Then there is an explicit surjective correspondence between a subset $\Lambda_A\subseteq\Lambda$ and the Euclidean space $\fr{a}^*$, defined as the dual space of the Lie algebra $\fr{a}$ induced by the Lie group $A$ of the abelian part of the Iwasawa decomposition of $G$.
For any stationary $k$ satisfying $\Bbbk \in L^2(G)$, we have $\supp \mu_k \subseteq \Lambda_A$ and $\mu_k$ is absolutely continuous with respect to the Lebesgue measure on $\fr{a}^*$.
\end{proposition}

\begin{proof}
Follows from \Cref{thm:k_helg}, which is presented subsequently.
\end{proof}

This tells us what set to integrate over: to ease notation, where unambiguous we henceforth use $\lambda$ to refer simultaneously to irreducible unitary representations $\lambda\in\Lambda$, and Lie algebra dual elements $\lambda \in \fr{a}^*$.
We proceed to understand how to carry out the necessary integration.

\subsection{Pointwise Kernel Evaluation}
\label{sec:computation:pointwise}

\begin{figure}
\begin{subfigure}{0.33\textwidth}
\begin{tikzpicture}
\path[use as bounding box] (-2.625,-1.5) rectangle (2.125,1.5);
\node at (0,0) {\includegraphics[scale=0.25]{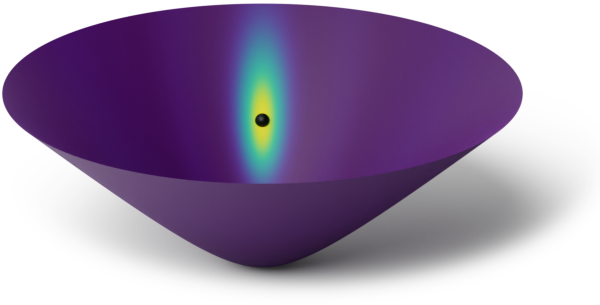}};
\end{tikzpicture}
\caption{Hyperbolic space}
\end{subfigure}
\begin{subfigure}{0.66\textwidth}
\begin{tikzpicture}
\path[use as bounding box] (-2.625,-1.5) rectangle (6.875,1.5);
\node at (0,0) {\includegraphics[scale=0.25]{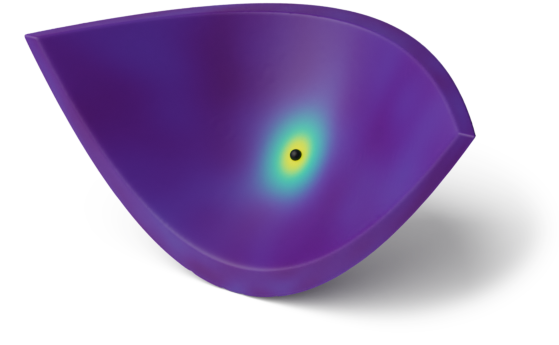}};
\node at (4.5,0) {\includegraphics[scale=0.25]{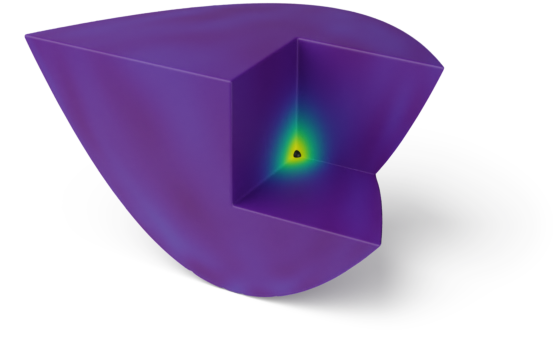}};
\end{tikzpicture}
\caption{Symmetric positive-definite matrices}
\end{subfigure}
\caption{Values of the heat kernel $k(\bullet,\.)$ on $\mathbb{H}_2$ and $\SPD(2)$.}
\label{fig:kernel-values}
\end{figure}

We now leverage the Iwasawa decomposition to reduce the abstract integral of \Cref{thm:stationary_symmetric}, which is supported on the set of irreducible unitary representations $\Lambda$, to a substantially more concrete integral over the space $\fr{a}^*$ introduced in \Cref{thm:iwasawa_support}.
To do this, observe that the Lie algebra $\fr{a}$ is a vector space and hence can be equipped with an inner product.
This inner product can be carried over to $\fr{a}^*$, which can then be used to equip $\fr{a}^*$ with a Lebesgue measure.
We can explicitly describe the integral of interest using this measure.

\begin{proposition} \label{thm:k_helg}
Let $\Bbbk \in L^2(G)$.
Then there is a non-negative $a^{(\lambda)} \in L^2(\fr{a}^*, \abs{c(\lambda)}^{-2} \d \lambda)$, where the $L^2$-weight function is defined in terms of Harish-Chandra's $c$-function, such that
\[
k(g_1 \bdot H, g_2 \bdot H)
=
\int_{\Lambda}
\pi^{(\lambda)}(g_2^{-1} \bdot g_1) \d\mu_k(\lambda)
=
\int_{\fr{a}^*} a^{(\lambda)}
\pi^{(\lambda)}(g_2^{-1} \bdot g_1) \abs{c(\lambda)}^{-2} \d\lambda
\]
where integration is defined with respect to the Lebesgue measure $\d\lambda$, and $\pi^{(\lambda)}$ in the latter case is the zonal spherical function obtained via the bijective correspondence of \Cref{thm:iwasawa_support}.
\end{proposition}

\begin{proof}
This follows largely by specializing ideas of \textcite{ggahelgason2000} to the setting at hand: full details are given in \Cref{apdx:spherical_fourier_transform}.
\end{proof}

This substantially simplifies the matter: our kernel is now expressed as an ordinary integral over a Euclidean space.
All that remains is to understand how to compute zonal spherical functions $\pi^{(\lambda)}$.
One can show these functions are expressed by explicit integrals.

\begin{result}
\label{thm:harish-chandra}
For $\lambda\in\fr{a}^*$, we have that
\[ \label{eqn:sph_func}
\pi^{(\lambda)}(g) = \int_H e^{(i\lambda+\rho)^{\top}a(h \bdot g)} \d \mu_H(h),
\]
where $a(g)\in\fr{a}$ satisfy $A(g) = \exp a(g)$, the term $\exp: \fr{g} \-> G$ is the exponential map, $\rho\in \fr{a}^*$ is an explicit vector given in terms of representation-theoretic quantities of $G$ which for most cases can be looked up in the literature,\footnote{More precisely, $\rho = \frac{1}{2}\sum_{\alpha\in \Sigma^+} m_\alpha \alpha$ is the half-sum of all positive roots $\alpha$ weighted by the dimensions $m_\alpha$ of the corresponding root spaces $\fr{g}_\alpha$---see the definitions of \emph{roots} and \emph{root spaces} as well as the appropriate references in \Cref{apdx:noncompact}.} and $\mu_H$ is the normalized Haar measure on $H$.
\end{result}

\begin{proof}
\textcite[Chapter IV, Theorem 4.3]{ggahelgason2000}.
\end{proof}

This completes the description, and establishes a way to evaluate the kernel pointwise. 
An illustration of kernel evaluations, for the heat kernel defined in \Cref{sec:heat_matern} and computed using this method, can be seen in \Cref{fig:kernel-values}.
For every $g\in G$ and $\lambda\in \fr{a}^*$, using Monte Carlo we have the following approximation 
\[
\label{eqn:sym_func_approx}
&\pi^{(\lambda)}(g) \approx \frac{1}{L}\sum_{l=1}^{L} e^{(i\lambda+\rho)^{\top}a(h_l \bdot g)} & h_l \~ \mu_H
\]
where we note that random sampling from the Haar measure on a compact group $H$ is generally tractable \cite{mezzadri07}.
From this, we have the random Fourier feature \cite{rahimi08} expansion
\[ 
\label{eqn:approx_evaluation}
\Bbbk(g) &\approx \frac{\sigma^2}{L} \sum_{l=1}^L e^{(i\lambda_l+\rho)^{\top}a(h_l \bdot g)}
&
\lambda_l &\~ \frac{1}{\sigma^2} \mu_k
&
h_l &\~ \mu_H
\]
where $\Bbbk(e) = \sigma^2$, and $e$ is the identity of $G$.
We now quantify its approximation error.

\begin{restatable}{proposition}{KernelEstimatorVar} \label{thm:kernel_estimator_var}
The standard deviation of the estimator \Cref{eqn:approx_evaluation} is uniformly bounded by $\frac{\sigma^2}{\sqrt{L}}$.
\end{restatable}
\begin{proof}
Direct computation using the properties of $\pi^{(\lambda)}$, see \Cref{apdx:proofs}.
\end{proof}

\subsection{Efficient Sampling}
\label{sec:computation:sampling}

The preceding expressions give us a way to evaluate the kernel pointwise.
They do not, however, immediately give a way to efficiently sample from the prior.
To do this, we will employ a different way of expressing the zonal spherical functions.

\begin{result} \label{thm:spherical_symmetry}
For any $\lambda \in \fr{a}^*$ and $g_1, g_2 \in G$ we have
\[ \label{eqn:aara}
\pi^{(\lambda)}(g_2^{-1} \bdot g_1) =  \int_H e^{(i\lambda+\rho)^{\top}a(h\bdot g_1)}\overline{e^{(i\lambda+\rho)^{\top}a(h\bdot g_2)}} \d \mu_H(h)
\]
where $a$ has the same meaning as in~\Cref{thm:harish-chandra}. 
\end{result}
\begin{proof}
\textcite[Chapter III, Theorem 1.1]{helgason1994}.
\end{proof}

\begin{figure}
\begin{subfigure}{0.33\textwidth}
\begin{tikzpicture}
\path[use as bounding box] (-2.625,-1.5) rectangle (2.125,1.5);
\node at (0,0) {\includegraphics[scale=0.25]{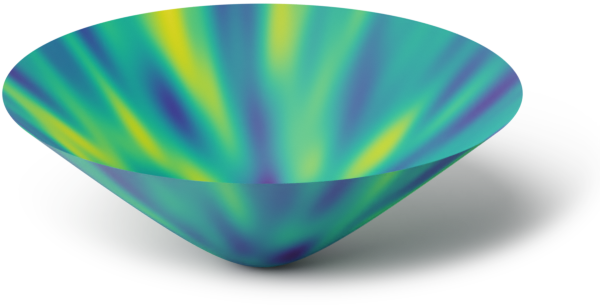}};
\end{tikzpicture}
\caption{Hyperbolic space}
\end{subfigure}
\begin{subfigure}{0.66\textwidth}
\begin{tikzpicture}
\path[use as bounding box] (-2.625,-1.5) rectangle (6.875,1.5);
\node at (0,0) {\includegraphics[scale=0.25]{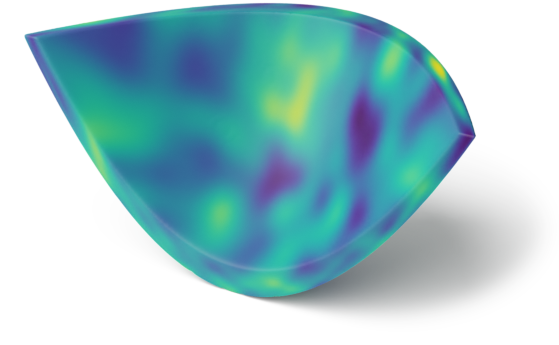}};
\node at (4.5,0) {\includegraphics[scale=0.25]{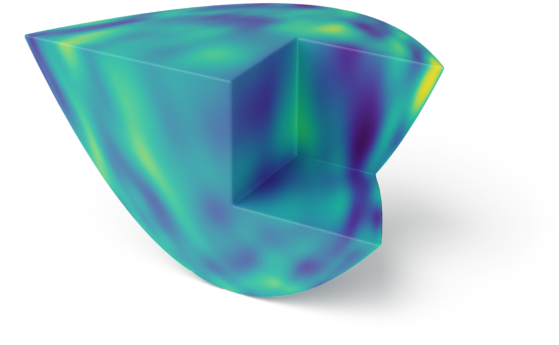}};
\end{tikzpicture}
\caption{Symmetric positive-definite matrices}
\end{subfigure}
\caption{Samples from a Gaussian process with heat kernel covariance on $\mathbb{H}_2$ and $\SPD(2)$.}
\label{fig:samples}
\end{figure}

Note the resemblance between this expression given in part I, specifically \textcite[equation~(35)]{part1}, from which one obtains generalized random phase Fourier features.
The key property of~\Cref{eqn:aara} is that the right-hand side can be viewed as an inner product.
Thanks to this property, the prior Gaussian process $f \~[GP](0, k)$ can be approximated as
\[ \label{eqn:rff_sym}
f(x) &\approx \frac{\sigma}{\sqrt{L}} \sum_{l=1}^L w_l e^{(i\lambda_l+\rho)^{\top}a(h_l \bdot x)}
&
\lambda_l &\~ \frac{1}{\sigma^2} \mu_k
&
h_l &\~ \mu_H
&
w_l \~[N](0, 1)
\]
where $\mu_H$ is the normalized Haar measure over $H$ and $\mu_k$ is the spectral measure.
We therefore obtain a way to use random Fourier features for efficient sampling: an illustration of such samples, for a Gaussian process with heat kernel covariance defined in \Cref{sec:heat_matern}, is given in \Cref{fig:samples}.
This is not the only use of \Cref{eqn:rff_sym}: the corresponding kernel approximation is by construction guaranteed to be positive semi-definite---unlike \Cref{eqn:approx_evaluation}.
This is given by
\[ 
\label{eqn:approx_evaluation_pd}
k(g_1 \bdot H, g_2 \bdot H) &\approx \frac{\sigma^2}{L} \sum_{l=1}^L e^{(i\lambda_l+\rho)^{\top} a(h_l\bdot g_1)}
\overline{e^{(i\lambda_l+\rho)^{\top} a(h_l\bdot g_2)}}
&
\lambda_l &\~ \frac{1}{\sigma^2} \mu_k
&
h_l &\~ \mu_H.
\]

One key difference from~\Cref{eqn:approx_evaluation}, as well as from the standard Euclidean case, is that the convergence rate of the approximation given by~\Cref{eqn:approx_evaluation_pd}, in the sense of the standard deviation of the estimated kernel, is not uniform, but rather only uniform on compact subsets.

\begin{restatable}{theorem}{RFFConvergence}
\label{thm:rff_convergence}
For any compact subset $U\subset G/H$ there is a constant $C_U > 0$ such that, for any $g_1 \bdot H, g_2 \bdot H \in U$, the right-hand side of~\Cref{eqn:approx_evaluation_pd} is an unbiased estimator of $k(g_1 \bdot H, g_2 \bdot H)$, whose standard deviation is bounded by $C_U\frac{\sigma^2}{\sqrt{L}}$. 
The constant $C_U$, however, does not admit an upper bound: for any fixed $L$, the standard deviation on the full space $G/H$ is unbounded.
\end{restatable}
\begin{proof}
\Cref{apdx:proofs}. 
\end{proof}

\section{Heat and Matérn Kernels} \label{sec:heat_matern}

In the previous section we obtained general computational recipes for working with stationary Gaussian processes on non-compact symmetric spaces.
These depend on the specific kernel $k$ only through its spectral measure $\mu_k$, which by \Cref{thm:stationary_symmetric} in turn uniquely defines $k$.
We now study specific families of stationary Gaussian processes on non-compact symmetric spaces, infer their associated spectral measures $\mu_k$, and discuss general techniques for sampling~$\mu_k$.

To do this, we first need to define a Riemannian metric on $X = G/H$, a symmetric space satisfying \Cref{asm:symmetric_space_assumptions}.
We do so canonically as follows.
Let $\fr{g}$ be the Lie algebra of $G$.
Since $G$ is semi-simple, it follows there is a canonical bilinear form on $\fr{g}$, termed the \emph{Killing form}, which is non-degenerate.
By pushforward, the Killing form defines a bi-invariant pseudo-Riemannian metric on $G$.
For non-compact semi-simple groups, this metric is never Riemannian, since the Killing form has both positive and negative eigenvalues \cite[Theorem 2.14]{arvanitogeorgos2003}.
In spite of this, under \Cref{asm:symmetric_space_assumptions}, the metric which the quotient $G/H$ inherits from $G$ is actually \emph{Riemannian}.
We assume henceforth that $G/H$ is equipped with this metric.

We focus on Matérn kernels $k_{\nu, \kappa, \sigma^2}$, where $\sigma^2,\kappa,\nu$ are the amplitude, length scale, and smoothness parameters, respectively, as well as its infinite-smoothness limit, the squared exponential kernel $k_{\infty, \kappa, \sigma^2}$, which is also known as the heat, Gaussian, or RBF kernel.
These are the most popular Euclidean Gaussian process kernels in machine learning and statistics.

To define their analogs in the setting at hand, we adopt the heat-equation-based approach used in the compact case \cite{part1}.
This works as follows.
First, the Euclidean squared exponential kernel is reinterpreted as the fundamental solution of the heat equation.
Next, we generalize the heat equation from Euclidean spaces to Riemannian manifolds, by replacing the Euclidean Laplacian with the Laplace--Beltrami operator, getting the following.

\begin{definition}
\label{def:heat_kernel}
Let $X$ be a Riemannian manifold. 
Define the \emph{heat kernel} to be 
\[
k_{\infty,\kappa,\sigma^2}(x,x') = \frac{\sigma^2}{C_{\infty, \kappa}} \c{P}(\kappa^2/2,x,x')
\]
where $C_{\infty, \kappa}$ is the constant that ensures $\int_{X} k_{\infty,\kappa,\sigma^2}(x,x) \d x = \sigma^2$, while $\c{P}$ is the solution of
\[
\pd{\c{P}}{t}(t,x,x') &= \lap_{x} \c{P}(t,x,x')
&
\c{P}(0,x,x') &= \delta(x,x')
\]
where $\Delta_x$ is the Laplace--Beltrami operator acting on the $x$ variable, and $\delta(x,x')$ is the Dirac delta function, defined appropriately.
The differential equation above is precisely the \emph{heat equation}, which in the given setting admits a unique solution in $L^2(X)$ \cite{grigoryan2009}.
\end{definition}

This gives the appropriate analog of the squared exponential kernel.
The final step is to observe that Euclidean Matérn kernels can equivalently be defined as integrals of squared exponential kernels, via a kernel analogue of the gamma mixture of Gaussians expression for the $t$-distribution, which is the spectral measure of a Matérn kernel.
Using this observation, this expression can more generally be taken as the \emph{definition} of a Matérn kernel.

\begin{definition}
\label{def:matern_kernel}
Let $X$ be a Riemannian manifold. 
Define the \emph{Matérn kernel} to be
\[ \label{eqn:matern_integral_formula}
k_{\nu, \kappa, \sigma^2}(x,x')
=
\frac{\sigma^2}{C_{\nu, \kappa}}
\int_0^{\infty}
u^{\nu - 1 + n/2}
e^{-\frac{2 \nu}{\kappa^2} u}
\c{P}(u, x, x')
\d u
\]
where $C_{\nu, \kappa}$ is the constant that ensures $\int_{X} k_{\nu,\kappa,\sigma^2}(x,x) \d x = \sigma^2$ and $n = \dim X$.
\end{definition}

\begin{figure}
\begin{subfigure}{0.3\textwidth}
\begin{tikzpicture}
\path[use as bounding box] (-2.375,-1.25) rectangle (2.125,1.25);
\node at (0,0) {\includegraphics[scale=0.25]{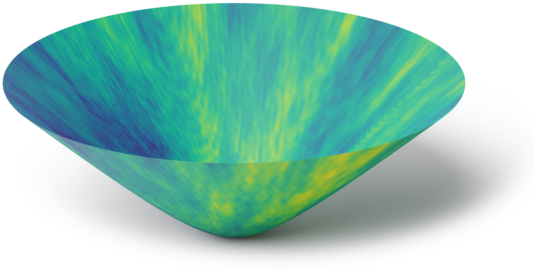}};
\end{tikzpicture}
\caption{$\nu = 0.5$}
\end{subfigure}
\begin{subfigure}{0.3\textwidth}
\begin{tikzpicture}
\path[use as bounding box] (-2.375,-1.25) rectangle (2.125,1.25);
\node at (0,0) {\includegraphics[scale=0.25]{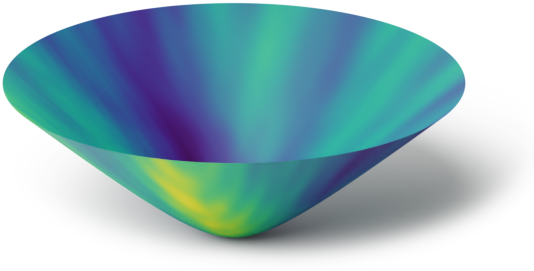}};
\end{tikzpicture}
\caption{$\nu = 1.5$}
\end{subfigure}
\begin{subfigure}{0.3\textwidth}
\begin{tikzpicture}
\path[use as bounding box] (-2.375,-1.25) rectangle (2.125,1.25);
\node at (0,0) {\includegraphics[scale=0.25]{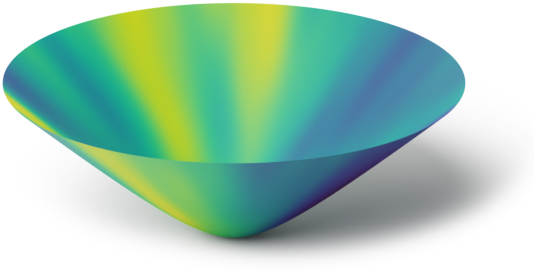}};
\end{tikzpicture}
\caption{$\nu = 2.5$}
\end{subfigure}
\caption{Samples from Matérn Gaussian processes on $\mathbb{H}_2$, with varying smoothness.}
\label{fig:matern-smoothness}
\end{figure}
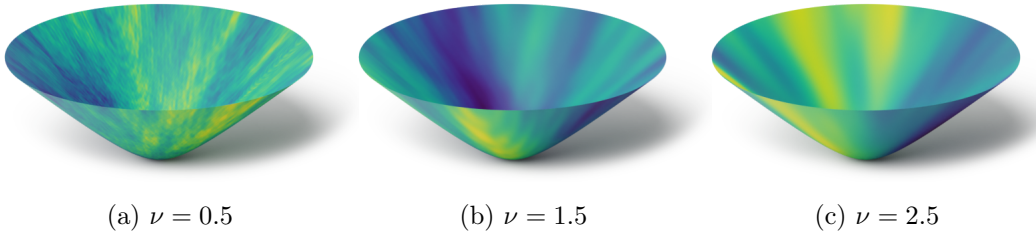

Note that both~\Cref{def:heat_kernel,def:matern_kernel} give the classical expressions in the Euclidean case.
We illustrate samples from Matérn Gaussian processes of varying smoothness in \Cref{fig:matern-smoothness}.

It turns out that, in contrast with the Euclidean and compact Riemannian cases, the above generalizations are not the only possibility: to understand this, we proceed to introduce a slightly different notion of a Matérn Gaussian process, by considering a modified Laplace operator.
We will prove both notions well-defined in~\Cref{sec:heat_matern:stat_and_sd}.

\subsection{Choice of Ordinary vs. Shifted Laplace--Beltrami Operator}

The preceding definitions of $k_{\infty,\kappa,\sigma^2}$ and $k_{\nu,\kappa,\sigma^2}$ depend on the manifold $X = G/H$ through the Laplace--Beltrami operator $\lap$.
One of the key ways in which non-compact symmetric spaces can differ from their compact counterparts is that the negated Laplacian $-\lap$ can be strongly elliptic: $\innerprod{-\lap f}{f}_{L^2(X)} \geq s \norm{f}_{L^2(X)}^2$ for some $s > 0$ and all $f \in L^2(X) \cap C^{\infty}(X)$.
The maximal $s$ such that the inequality holds is called the \emph{spectral gap}.\footnote{This notion of spectral gap is not to be confused with an alternative one, commonly used for compact Riemannian manifolds. This alternative notion is defined as the smallest \emph{non-zero} eigenvalue of $-\lap$.}

The spectral gap of the Laplace--Beltrami operator is always zero in the Euclidean and compact Riemannian cases.\footnotemark[\thefootnote]
The fact that it can be non-zero on non-compact symmetric spaces  can cause the differential equations such as the heat equation to behave in a qualitatively different manner: see, for instance, \textcite{vazquez2022, anker2023}.

In cases where a positive spectral gap $s > 0$ leads to undesired behavior, one can instead choose to work with the \emph{shifted Laplacian} $\lap+sI$ \cite{wildberger1995}, which is defined by adding a scalar multiple of the identity map $I$ to the Laplace--Beltrami operator~$\lap$.
Here, $s$ is the spectral gap of $\lap$.
Because of the strong ellipticity of $-\lap$, the negated shifted Laplacian $-\lap-sI$ is still positive semi-definite on $L^2(X)$---but its spectral gap vanishes, making it more similar to its Euclidean and compact Riemannian counterparts.
This operator is sometimes used in defining analogues of the heat equation \cite{maher2014} as well the analogues of wave equation \cite{helgason1992} and Helmholtz equation \cite{livshits2015} in the non-compact symmetric space setting.

Since heat and Matérn kernels can be defined just-as-easily with the shifted Laplacian as with the ordinary Laplacian, the results we develop can be used in both settings with minor modifications.
We will examine practical differences that result from working with the shifted Laplacian vs. the ordinary Laplacian in the sequel.
First, however, we study these kernels' basic properties, which will already reveal some of the consequences of this choice.

\subsection{Stationarity and Spectral Densities} \label{sec:heat_matern:stat_and_sd}

We now deduce a more explicit form for the kernels defined in \Cref{def:heat_kernel,def:matern_kernel}, proving them positive semi-definite along the way.
This entails computing the solution of the heat equation in the setting where $X = G / H$ is a symmetric space.
The standard way of doing this in the Euclidean case is by applying the Fourier transform, solving the resulting algebraic equation, then applying the inverse Fourier transform to map the solution back to the original space.
To proceed, we apply this argument adapted to the setting of non-compact symmetric spaces, using the \emph{spherical Fourier transform} of \textcite{helgason1994}.

\begin{proposition}
\label{prop:spectral_measure_of_kernels}
The heat and Matérn kernels on a non-compact symmetric space $X = G/H$ are positive semi-definite, stationary, and given by
\[ \label{eq:sym_space_matern_kernels}
k_{\nu, \kappa, \sigma^2}(g_1 \bdot H, g_2 \bdot H)
=
\int_{\fr{a}^*}
  \pi^{(\lambda)}(g_2^{-1} \bdot g_1)
  \d\mu_{\nu, \kappa, \sigma^2}(\lambda),
\]
where the respective spectral measures $\mu_{\nu,\kappa,\sigma^2}$ are
\[
\label{eq:sym_space_matern_spectral_measure}
\mu_{\nu, \kappa, \sigma^2}
&=
\begin{cases}
  \frac{\sigma^2}{C_{\infty, \kappa}'} e^{-\frac{\kappa^2}{2} \del{\norm{\lambda}^2 + \norm{\rho}^2}} |c(\lambda)|^{-2} \d \lambda 
  & 
  \nu = \infty
  \\
  \frac{\sigma^2}{C_{\nu, \kappa}'} \del{\frac{2 \nu}{\kappa^2} + \norm{\lambda}^2 + \norm{\rho}^2}^{-\nu-n/2} |c(\lambda)|^{-2} \d \lambda 
  & 
  \nu < \infty
\end{cases}
\]
where $\rho$ is defined in \Cref{thm:harish-chandra}, $c(\lambda)$ is Harish-Chandra's $c$-function, and $C'_{\nu,\kappa}$ are constants which ensure that $\int_{X} k_{\nu,\kappa,\sigma^2}(x,x) \d x = \sigma^2$ which in this case reduces to $k_{\nu,\kappa,\sigma^2}(x,x) = \sigma^2$.
Moreover, the kernels $\h{k}_{\nu,\kappa,\sigma^2}$ defined via the shifted Laplacian admit analogous expressions for their spectral measures $\h{\mu}_{\nu, \kappa, \sigma^2}$, but with $\norm{\rho}^2$ removed and with different constants~$C'_{\nu,\kappa}$.
\end{proposition}

\begin{proof}
A proof for the heat kernel ($\nu=\infty$) can be found in \textcite[Proposition 3.1]{Gangolli1968}, also see \Cref{apdx:spherical_fourier_transform} for a proof sketch.
From this, the Matérn kernel's form follows by applying its definition, changing the order of integration, and appropriate algebra, which works the same way as in the compact case---see \textcite{part1}.

The proof applies mutatis mutandis for $\h{k}_{\nu,\kappa,\sigma^2}$.
The fact that both $k_{\nu,\kappa,\sigma^2}$ and $\h{k}_{\nu,\kappa,\sigma^2}$ are positive semi-definite and stationary follows from their form in~\Cref{eq:sym_space_matern_kernels} and~\Cref{thm:stationary_symmetric}.
\end{proof}

We now state an important relation between the solution $\c{P}$ of the heat equation for the Laplace--Beltrami operator and the solution $\h{\c{P}}$ for its shifted counterpart which follows from the proof of \Cref{prop:spectral_measure_of_kernels}.

\begin{corollary}
We have
\[
\h{\c{P}}(t, x, x') = e^{t \norm{\rho}^2} \c{P}(t, x, x').
\]
It follows that $k_{\infty, \kappa, \sigma^2} = \h{k}_{\infty, \kappa, \sigma^2}$.
\end{corollary}

Note that this does \emph{not} imply the similar relation for $\nu < \infty$: the kernels $k_{\nu, \kappa, \sigma^2}$ and $\h{k}_{\nu, \kappa, \sigma^2}$ will typically still be different.

All that remains to compute the heat and Matérn kernels numerically using random Fourier features is to develop techniques for sampling from the explicit spectral measures.
The main obstacle that seemingly remains is Harish-Chandra's $c$-function, which produces densities that do not have a known form.
Before handling this, we verify that these kernels' hyperparameters correspond to the same regularity classes as they do in the Euclidean case.

\begin{restatable}{theorem}{SmoothnessNoncomp}
\label{thm:smoothness_noncomp}
For all $\nu, \kappa, \sigma^2 > 0$ and all $g_2 \in G$, the Matérn kernels $k_{\nu, \kappa, \sigma^2}(\., g_2 \bdot H)$ and $\h{k}_{\nu, \kappa, \sigma^2}(\., g_2 \bdot H)$ lie in the Sobolev space $H^{\nu+n/2}$ if $\nu < \infty$, and in all Sobolev spaces $H^{\alpha}$, $\alpha > 0$, if $\nu = \infty$. Thus, they are continuous and possess continuous derivatives of all integer orders strictly less than $\nu$.
\end{restatable}

\begin{proof}
The first part follows from a certain spectral characterization of $H^{\nu+n/2}$.
The second part follows from an appropriate Sobolev embedding theorem.
See \Cref{apdx:proofs}.
\end{proof}

\subsection{General Computation Techniques}

We now study how to sample from the spectral measures $\mu_{\nu, \kappa, \sigma^2}$, which is all that remains for applying the techniques of \Cref{sec:computation}.
Here, we present general techniques that do not depend on the space being studied: \Cref{sec:spd_hyp} gives specialized techniques for hyperbolic space and the space of positive definite matrices, leveraging their special structure to improve efficiency compared to the techniques we now describe.
We work with the ordinary Laplacian: the same techniques also apply to the shifted Laplacian, provided that one sets $\rho = 0$ and redefines the resulting normalizing constants appropriately.
Observe that the densities
\[
p_{\nu,\kappa}(\lambda)
\propto
\begin{cases}
e^{-\frac{\kappa^2}{2}\del{\norm{\lambda}^2 + \norm{\rho}^2}} & \nu = \infty
\\
\del{\frac{2 \nu}{\kappa^2} + \norm{\lambda}^2 + \norm{\rho}^2}^{-\nu-n/2} & \nu < \infty
\end{cases}
\]
are precisely the usual multivariate Gaussian and (generalized) $t$-densities.
Using these densities, the respective spectral measures can be expressed as 
\[
\mu_{\nu, \kappa, \sigma^2} \propto \sigma^2 p_{\nu,\kappa}(\lambda) |c(\lambda)|^{-2} \d \lambda
.
\]
The analogs $\h{p}_{\nu,\kappa}$ of $p_{\nu,\kappa}$ for the shifted Laplacian and $\h{\mu}_{\nu, \kappa, \sigma^2}$ are obtained by setting $\norm{\rho}^2 = 0$.

We can therefore apply importance sampling to obtain
\[
\label{eqn:kernel_imp_sampling_approx}
\Bbbk(g)
&\approx
\frac{\sigma^2}{C_{\nu, \kappa}'' L}
\sum_{l=1}^L
\abs{c(\lambda_l)}^{-2}
e^{(i\lambda_l+\rho)^{\top}a(h_l \bdot g)}
&
\lambda_l &\~ p_{\nu,\kappa}
&
h_l &\~ \mu_H
&
&
\\
\label{eqn:rff_with_c_func}
f(g\bdot H)
&\approx
\frac{\sigma^2}{\sqrt{C_{\nu, \kappa}'' L}}
\sum_{l=1}^L
w_l
|c(\lambda_l)|^{-1}
e^{(i\lambda_l+\rho)^{\top}
a(h_l \bdot g)}
&
\lambda_l &\~ p_{\nu,\kappa}
&
h_l &\~ \mu_H
&
w_l &\~[N](0, 1)
\]
where $\mu_H$ is the normalized Haar measure on $H$, and
$C_{\nu, \kappa}''$ are normalization constants chosen to ensure the approximation has variance $\sigma^2$.

Note that the kernel approximation corresponding to the right-hand side of~\Cref{eqn:rff_with_c_func}---the kernel approximation which is necessarily positive semi-definite---is given by
\[
\label{eqn:rff_with_c_func_kern}
k(g_1 \bdot H, g_2 \bdot H) &\approx
\frac{\sigma^2}{C_{\nu, \kappa}'' L} \sum_{l=1}^L
\abs{c(\lambda_l)}^{-2}
e^{(i\lambda_l+\rho)^{\top} a(h_l\bdot g_1)}
\overline{e^{(i\lambda_l+\rho)^{\top} a(h_l\bdot g_2)}}
&
&\begin{aligned}
\lambda_l &\~ p_{\nu,\kappa}
\\
h_l &\~ \mu_H
.
\end{aligned}
\]

Since these estimators are based on importance sampling, it is easy to see that they are less efficient compared to direct Monte Carlo sampling.
We will demonstrate in~\Cref{sec:spd_hyp} that for certain spaces it is possible to instead derive efficient rejection samplers, as long as one is willing to work with space-specific techniques.

\section{Examples} \label{sec:spd_hyp}

We now specialize the techniques developed to the hyperbolic space $\mathbb{H}_n$ and the space of symmetric positive-definite matrices $\SPD(d)$---arguably the two most important examples of non-compact symmetric spaces---and propose efficient approximations for kernels on those spaces, leveraging their specific structure.
The key idea is that densities of the measures~$\mu_{\nu, \kappa, \sigma^2}$, and respectively $\h{\mu}_{\nu, \kappa, \sigma^2}$, can be upper-bounded by tractable densities, allowing efficient rejection sampling.
For $\bb{H}_n$ these are the simple mixtures of chi or beta-prime distributions, while for $\SPD(d)$ these are the densities arising as the distributions of eigenvalues of certain random matrices.
Moreover, in this section, we will also discuss the Riemannian metric used on $\SPD(d)$, and handle the fact that this space does not satisfy the formal assumptions of~\Cref{sec:computation} by representing it as a suitable product of spaces that do or are otherwise well-behaved, following the strategy discussed alongside \Cref{asm:symmetric_space_assumptions}.

\subsection{Hyperbolic Space}
\label{sec:hyperbolic_space}

\begin{figure}
\begin{subfigure}{0.24\textwidth}
\begin{tikzpicture}
\path[use as bounding box] (-1.875,-1) rectangle (1.75,1);
\node at (0,0) {\includegraphics[scale=0.25]{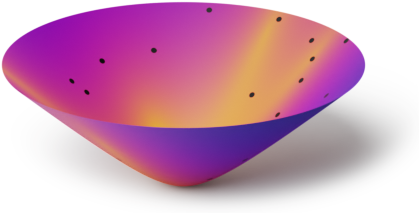}};
\end{tikzpicture}
\caption{Ground truth}
\end{subfigure}
\begin{subfigure}{0.24\textwidth}
\begin{tikzpicture}
\path[use as bounding box] (-1.875,-1) rectangle (1.75,1);
\node at (0,0) {\includegraphics[scale=0.25]{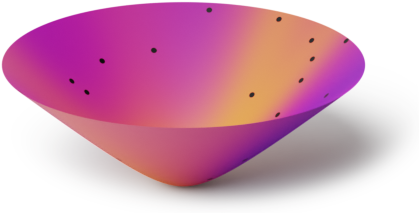}};
\end{tikzpicture}
\caption{Mean}
\end{subfigure}
\begin{subfigure}{0.24\textwidth}
\begin{tikzpicture}
\path[use as bounding box] (-1.875,-1) rectangle (1.75,1);
\node at (0,0) {\includegraphics[scale=0.25]{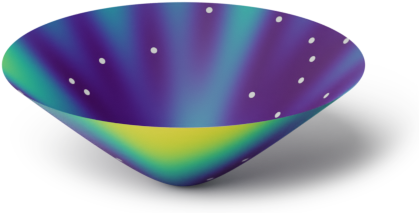}};
\end{tikzpicture}
\caption{Standard dev.}
\end{subfigure}
\begin{subfigure}{0.24\textwidth}
\begin{tikzpicture}
\path[use as bounding box] (-1.875,-1) rectangle (1.75,1);
\node at (0,0) {\includegraphics[scale=0.25]{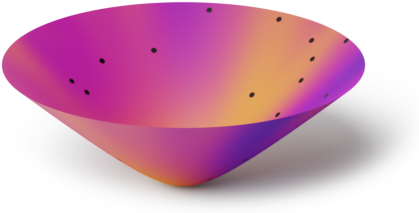}};
\end{tikzpicture}
\caption{Posterior sample}
\end{subfigure}
\caption{Gaussian process regression on hyperbolic space $\mathbb{H}_2$ driven by the heat kernel.}
\label{fig:hyperbolic-posterior}
\end{figure}
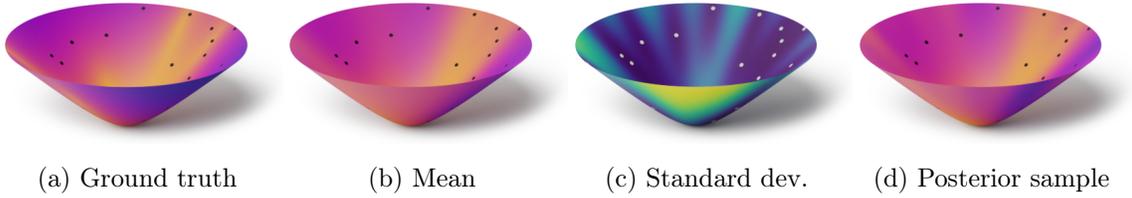

Hyperbolic space $\bb{H}_n$ is defined as the unique, up to an isometry, simply connected $n$-dimensional Riemannian manifold with sectional curvature $-1$.
More concretely, it is usually represented by \emph{models} which are concrete manifolds that are isometric to the general definition.
For instance, the \emph{Poincaré ball model} identifies $\bb{H}_n$ with the open unit ball $\f{B}(n) = \cbr{\v{x} \in \R^n : \norm{\v{x}} < 1} \subseteq \R^n$ endowed with the non-Euclidean Riemannian metric
\[
g_{\v{x}} (\v{u}_{\v{x}}, \v{v}_{\v{x}})
=
\frac{4\innerprod{\v{u}_{\v{x}}}{\v{v}_{\v{x}}}}{\big(1-\norm{\v{x}}^2\big){\vphantom{\norm{x}}}^2},
\]
where $\v{x} \in \f{B}(n)$ and $\v{u}_{\v{x}},\v{v}_{\v{x}}$ are tangent vectors \cite[Corollary~2.4.13]{wolf2011}.
Alternatively, $\bb{H}_n$ can be represented by the \emph{hyperboloid model}.
To construct this model, one starts with generalized Minkowski space, defined as $\R^{n+1}$ with the bilinear form
\[ \label{eqn:minkowski_product}
\innerprod{\v{v}}{\v{u}}_M = -v_0 u_0 + v_1 u_1 + \ldots + v_n u_n
\]
in place of the standard inner product.
This is a pseudo-Euclidean space, and hence a pseudo-Riemannian manifold, the $n=3$ version of which models spacetime in special relativity.
One then considers the unit sphere $\cbr{\v{v}\in\R^{n+1} : \innerprod{\v{v}}{\v{v}}_M=1}$, which is a two-sheeted hyperboloid.
Hyperbolic space is realized as its forward sheet, that corresponds to $v_0>0$, which is a Riemannian manifold.
We show a simple Gaussian process regression model on $\bb{H}_n$, computed using the techniques of this section, in \Cref{fig:hyperbolic-posterior}.

The hyperboloid model helps identify that $\bb{H}_n$ is a homogeneous space.
Let $\SL(1 + n)$ be the group of $(1+n) \x (1+n)$ matrices with unit determinant.
Define the \emph{proper Lorentz group} $\SO(1, n)$ as the subgroup of $\SL(1 + n)$ that preserves $\innerprod{\.}{\.}_M$, namely
\[
\SO(1, n)
=
\cbr{\m{M} \in \SL(1 + n) : \innerprod{\m{M}\v{v}}{\m{M}\v{u}}_M = \innerprod{\v{v}}{\v{u}}_M, \forall \v{v}, \v{u} \in \R^{n+1}}.
\]
The proper Lorentz group's action on $\bb{H}_n$ is simply the restriction onto the unit sphere of the natural action of $\SL(1+n)$ on $\R^{n+1}$.
The group $\SO(1, n)$ is not connected and has two connected components.
Let $\SO_0(1, n)$ be the connected component that contains the identity matrix.
It consists \cite{lester1993} of $\m{M} \in \SO(1, n)$ such that
\[
\m{M}
&=
\begin{pmatrix}
a & \v{b}^\top \\
\v{c} & \m{D}
\end{pmatrix}
&
a &> 0
&
\v{b},\v{c} &\in \R^n
&
\det \m{D} &> 0.
\]
Let $\v{x}_0 = (0, 1/\sqrt{n},\ldots,1/\sqrt{n})$.
Then the subgroup of $\SO_0(1,n)$ which preserves $\v{x}_0$, in the sense  $\m{M} \v{x}_0 = \v{x}_0$, will be the group $H \subseteq \SO_0(1,n)$ that is isomorphic to the group $\SO(n)$ of orthogonal matrices with unit determinant.
With these definitions, we have the~following.

\begin{result}
The quotient manifold $\SO_0(1, n) / \SO(n)$ is a symmetric space satisfying~\Cref{asm:symmetric_space_assumptions}.
Moreover, when equipped with the canonical metric induced by the Killing form, this space is isometric to $\bb{H}_n$.
\end{result}

\begin{proof}
See \textcite[Theorem~8.12.2(iii.b')]{wolf2011}.
\end{proof}

\subsubsection{Explicit Formulas for Spherical Functions and Spectral Densities}

On $\bb{H}_n$, the set $\fr{a}^*$ that parametrizes the irreducible unitary representations of interest corresponding to $\bb{H}_n$ is one-dimensional: $\fr{a}^* = \R$. 
Moreover, $\rho = \frac{n-1}{2}$.
The integral form of the zonal spherical functions on $\bb{H}_n$ ostensibly involves integration over $\SO(n)$: however, this integral actually simplifies further.
The basic reason for this is that, for $a$ from~\Cref{thm:harish-chandra}, there are distinct values $\m{V}, \m{U} \in H \subseteq \SO_0(1, n)$ for which $a(\m{V} \m{M}) = a(\m{U} \m{M})$ for all $\m{M} \in \SO_0(1, n)$, which allows one to instead integrate over a smaller set.
We formally state the underlying symmetry in terms of the diagonal part $A(\m{M})$ of the Iwasawa decomposition of elements $\m{M} \in \SO_0(1, n)$---recall that $A$ and $a$ are connected by $A(\m{M}) = \exp a(\m{M})$.

\begin{restatable}{lemma}{SnIntegrationLemma}
We have $A(\m{U} \m{M}) = A(\m{M})$ for all $\m{M} \in \SO_0(1, n)$ and all $\m{U} \in H \subseteq \SO_0(1,n)$ of the form
\[
\m{U}
&=
\begin{pmatrix}
1 & 0 & \\
0 & 1 & \\
  &   & \tl{\m{U}} \\
\end{pmatrix}
&
\tl{\m{U}} &\in \SO(n-1).
\]
\end{restatable}

\begin{proof}
This follows from the Iwasawa decomposition for $\SO_0(1, n)$, see~\Cref{apdx:proofs}.
\end{proof}

Consequently, the same holds for $a(\.)$, making it possible to use this symmetry to replace integration over $\SO(n)$ with integration over $\SO(n)/\SO(n-1) \isom \bb{S}_n$, the $n$-dimensional sphere.
In the Poincaré disk model, the resulting zonal spherical functions admit a particularly simple form \cite{limin1994}.
Here, the sphere $\bb{S}_n$ appears as boundary of the unit ball $B(n)$ that models hyperbolic space, and the zonal spherical functions are given by
\[
\pi^{(\lambda)}(\v{x})
&=
\int_{\bb{S}_n}
\del{\frac{1-\norm{\v{x}}^2}{\norm{\v{x}-\v{b}}^2}}^{i|\lambda|+(n-1)/2}
\d\mu_{\bb{S}_n}(\v{b})
&
\v{x} &\in B(n) \subseteq \R^n
&
\lambda \in \R
\]
where $\mu_{\bb{S}_n}$ is the Haar measure on the $n$-dimensional sphere.
For $\bb{H}_n$, Harish-Chandra's $c$-function depends on the parity of the space's dimension.
It is given by \textcite{limin1994} as
\[
\label{eqn:hyp_c}
|c(\lambda)|^{-2} &\propto 
\begin{cases}
\displaystyle 
|\lambda|\tanh(\pi|\lambda|)\prod_{j=2}^{m} \del{|\lambda|^2 + \frac{(2j-3)^2}{4}} 
\quad
&
n=2m
\\
\displaystyle 
\prod_{j=0}^{m-1} \del{|\lambda|^2 + j^2} 
\quad
&
n=2m+1.
\end{cases}
\]

With this, applying the general theory of \Cref{sec:computation} allows one to approximate heat and Matérn kernels numerically, and perform efficient sampling of the respective Gaussian processes.

\subsubsection{Efficient Sampling From Spectral Densities}

In \Cref{sec:computation}, we showed that the problem of computing heat and Matérn kernels can ultimately be reduced to the problem of sampling from their respective spectral measures.
We now develop specific techniques for doing so on $\bb{H}_n$.
We concentrate on the ordinary Laplace--Beltrami operator, remarking on how to apply the same techniques to the shifted Laplacian at the end of this section.
First, let us expand the polynomials in \Cref{eqn:hyp_c}:
\[
|c(\lambda)|^{-2}
\propto
\begin{cases}
\displaystyle
\tanh(\pi|\lambda|) \sum_{j=0}^{m-1} a_{n, j} \abs{\lambda}^{2j+1}
\quad 
&
n = 2m
\\
\displaystyle
\sum_{j=0}^{m} a_{n,j} \abs{\lambda}^{2j} 
\quad
&
n = 2m+1
\end{cases}
\]
where $a_{n,j}$ are the coefficients of the respective polynomials.
Using this, and $\rho = (n-1)/2$, we can express heat kernels' spectral measure as
\[
\mu_{\infty,\kappa,\sigma^2} \propto
\begin{cases}
\displaystyle
\tanh(\pi |\lambda|) \sum_{j=0}^{m-1}
a_{n, j} \abs{\lambda}^{2j + 1}
e^{-\frac{\kappa^2}{2}\abs{\lambda}^2}
\d \lambda
\quad
&
n = 2m
\\
\displaystyle
\sum_{j=0}^{m}
a_{n, j} \abs{\lambda}^{2j}
e^{-\frac{\kappa^2}{2}\abs{\lambda}^2}
\d \lambda
\quad 
&
n = 2m + 1.
\end{cases}
\]
Note that we dropped the $\del{\frac{n-1}{2}}^2$ term from the exponent: we can do this because it only changes the normalizing constant.
This density takes the form of a Gaussian times a polynomial, which is an analytically tractable form. 
Since the spectral densities and spherical functions depend only on the absolute value of $\lambda$, we may also restrict support of $\lambda$ onto~$[0,\infty)$.

\begin{restatable}{proposition}{HeatHypSamp}
\label{thm:hyperbolic_sampling_heat}
For any $m \in \N$, and any coefficients $\alpha_j \geq 0$, the density
\[
p_{\infty,\kappa}(\lambda)
&\propto
\sum_{j=0}^{m}
\alpha_j \lambda^{j}
e^{-\frac{\kappa^2}{2}\lambda^2}
&
\lambda &\in \R_{>0}
\]
is a mixture of scaled chi-distributed random variables.
Specifically, we have
\[
p_{\infty,\kappa}(\lambda) &= \sum_{j=0}^m c_j p_{y_j}(\lambda)
&
c_j &= \frac{\alpha_j/\beta_j}{\sum_{j=0}^m \alpha_j/\beta_j}
&
\beta_j
&=
2^{\frac{1-j}{2}}
\Gamma\del{\frac{j+1}{2}}^{-1}
\kappa 
\]
where $\Gamma(\.)$ is the gamma function, and $p_{y_j}(\lambda)$ is the density of the random variable $y_j = x_j/\kappa$ where $x_j \sim \chi_{j+1}$ is a chi-distributed random variable with $j+1$ degrees of freedom.
\end{restatable}
\begin{proof}
Direct computation, see \Cref{apdx:proofs}.
\end{proof}

For odd $n$, this immediately yields a way to sample exactly from the spectral density.
For even $n$, this is \emph{almost} the heat kernel's spectral density: once constants are dropped: all that remains is the hyperbolic tangent term.
Since $\tanh(x) < 1$, we can devise an efficient rejection sampler.
Before stating the algorithm, we extend this idea to the Matérn class. 
Using~\Cref{eq:sym_space_matern_spectral_measure} we can express Matérn kernel's spectral density as
\[
\mu_{\nu,\kappa,\sigma^2} \propto
\begin{cases}
\displaystyle
\tanh(\pi \abs{\lambda}) \sum_{j=0}^{m-1}
a_{n, j} \abs{\lambda}^{2j + 1}
\del{\frac{2\nu}{\kappa^2} + |\lambda|^2+\abs{\frac{n-1}{2}}^2}^{\mathrlap{-\nu-n/2}}
\qquad
\d \lambda
&
n = 2m
\\
\displaystyle
\sum_{j=0}^{m}
a_{n, j} \abs{\lambda}^{2j}
\del{\frac{2\nu}{\kappa^2} + |\lambda|^2+\abs{\frac{n-1}{2}}^2}^{\mathrlap{-\nu-n/2}}
\qquad
\d \lambda
&
n = 2m + 1.
\end{cases}
\]
In this case the respective part of the spectral density, excluding the hyperbolic tangent, is again tractable.

\begin{restatable}{proposition}{MaternHypSamp}
\label{thm:hyperbolic_sampling_matern}
For any $m \in \N$, $m \leq n-1$, and any coefficients $\alpha_j \geq 0$, the density 
\[
p_{\nu,\kappa}(\lambda)
&\propto
\sum_{j=0}^m
\alpha_j
\lambda^{j}
\del{
\frac{2\nu}{\kappa^2}
+
\lambda^2
+
\del{\frac{n-1}{2}}^2
}^{-\nu-\frac{n}{2}}
&
\lambda &\in \R_{>0}
\]
is a mixture of transformed beta-prime distributed random variables.
Specifically, we have 
\[
p_{\nu,\kappa}(\lambda) &=
\sum_{j=0}^m
c_j
p_{y_j}(\lambda)
&
c_j &= \frac{\alpha_j/\beta_j}{\sum_{j=0}^m \alpha_j/\beta_j}
&
\beta_j &= 2 \f{B}\del{\frac{j\phantom{*}\mathclap{+}\phantom{*}1}{2},\nu\phantom{*}\mathclap{+}\phantom{*}\frac{n\phantom{*}\mathclap{-}\phantom{*}j\phantom{*}\mathclap{-}\phantom{*}1}{2}}^{\phantom{**}\mathllap{-1}} \gamma^{\nu+\frac{n-j-1}{2}}
\]
where $\f{B}$ is the beta function, $\gamma = \frac{2\nu}{\kappa^2} + \del{\frac{n-1}{2}}^2$, and $p_{x_j}$ is the density of the random variable $y_j = \sqrt{\gamma x_j}$ where $x_j \sim \mathrm{Beta}'(\frac{j+1}{2}, \nu+\frac{n-j-1}{2})$ is beta-prime distributed.
\end{restatable}
\begin{proof}
Direct computation, see \Cref{apdx:proofs}.
\end{proof}

Using this, for both heat and Matérn kernels, we obtain a rejection-sampling-based strategy for sampling from the respective spectral measure.

\begin{algorithm}
Rejection sampler for $\mu_{\nu,\kappa,\sigma^2}$ on $\bb{H}_n$.
\1 Compute $a_j$ by expanding \Cref{eqn:hyp_c}.
\2 Sample the mixture component index $j$ with probability $c_j$ from \Cref{thm:hyperbolic_sampling_heat} or \Cref{thm:hyperbolic_sampling_matern}, for the heat or Matérn cases, respectively.
\3 Sample $\lambda$ from $p_{Y_j}$ where $p_{Y_j}$ is given in \Cref{thm:hyperbolic_sampling_heat} or \Cref{thm:hyperbolic_sampling_matern}, for the heat or Matérn cases, respectively.
\4 If $n$ is even, reject $\lambda$ with probability $1-\tanh(\pi\lambda)$ returning to step 2, otherwise proceed.
\5 Uniformly sample a sign $s \in \cbr{+1, -1}$ and output $s \lambda$.
\0
\end{algorithm}

\begin{remark}
Setting $\gamma = \frac{2 \nu}{\kappa}$ instead of $\gamma = \frac{2\nu}{\kappa^2} + \del{\frac{n-1}{2}}^2$ results in an algorithm for sampling the measure $\h{\mu}_{\nu, \kappa, \sigma^2}$ that corresponds to the shifted Laplacian, instead of the measure $\mu_{\nu, \kappa, \sigma^2}$.
\end{remark}

This completes our development: we can now compute heat and Matérn kernels, and perform efficient sampling from the prior.
Note that, here, $\lambda$ is one-dimensional, which follows from $\bb{H}_n$ being a rank-$1$ symmetric space.
This is a key reason for the relative simplicity of these results.

\subsubsection{Closed Form Expressions and Recurrence Relations for Heat Kernels}

It is known \cite{jaquier2020, grigoryan1998} that in odd dimensions heat kernels on hyperbolic spaces admit closed-form expressions, while in even dimensions the general formulas simplify to certain one-dimensional integrals.
For $x, x' \in \bb{H}_n$ denote the geodesic distance between $x$ and $x'$ by $\rho = \f{dist}_{\bb{H}_n}(x, x')$.
Then, in two and three dimensions, we have
\[ \label{eqn:hyp_heat}
k^{\bb{H}_2}_{\infty, \kappa, \sigma^2}(x, x')
&=
\frac{\sigma^2}{C_{\infty, \kappa}}
\int_{\rho}^{\infty}
\frac{s \exp\del{-s^2/(2 \kappa^2)}}{\del{\cosh(s) - \cosh(\rho)}^{1/2}} \d s
\\
k^{\bb{H}_3}_{\infty, \kappa, \sigma^2}(x, x')
&=
\frac{\sigma^2}{C_{\infty, \kappa}}
\frac{\rho}{\sinh{\rho}} e^{-\frac{\rho^2}{2 \kappa^2}}
\]
and other cases may be obtained by the following recurrence relation, known as Millson's formula, for $d > 3$:
\[
k^{\bb{H}_d}_{\infty, \kappa, \sigma^2}(x, x')
=
-
\frac{\sigma^2}{C_{\infty, \kappa} \sinh{\rho}}
\frac{\partial}{\partial \rho}
k^{\bb{H}_{d-2}}_{\infty, \kappa, \sigma^2}(x, x').
\]
Note that, unfortunately, for even dimensions the obvious Monte Carlo approximations of these integrals are not guaranteed to be positive semi-definite.

\subsection{The Space of Symmetric Positive-definite Matrices \texorpdfstring{$\SPD(d)$}{SPD(d)}} \label{sec:spd}

Symmetric positive-definite matrices arise in many real-world applications~\cite{Arnaudon2013,Dong2015, jaquier2020, jayasumana2013}.
We denote the set of such matrices of size $d\x d$ by $\SPD(d)$.
This is an open subset \cite[Section~1.1.2]{terras2016} of the $\frac{d(d+1)}{2}$-dimensional vector space of symmetric matrices, from which it inherits the structure of a smooth manifold.
There are different Riemannian metrics on $\SPD(d)$ \cite{jayasumana2013,lin2019,dryden2009}: we will concentrate on the \emph{affine-invariant metric}, which makes $\SPD(d)$ into a symmetric space of negative curvature.

\begin{figure}
\begin{subfigure}{0.3\textwidth}
\begin{tikzpicture}
\path[use as bounding box] (-2.375,-1.5) rectangle (2,4.75);
\node at (0,3.125) {\includegraphics[scale=0.25]{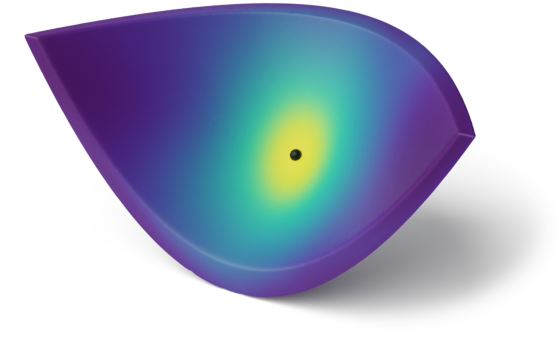}};
\node at (0,0) {\includegraphics[scale=0.25]{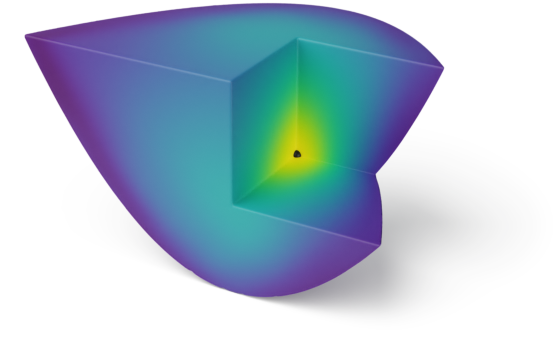}};
\end{tikzpicture}
\caption{$\kappa = 0.1$}
\end{subfigure}
\begin{subfigure}{0.3\textwidth}
\begin{tikzpicture}
\path[use as bounding box] (-2.375,-1.5) rectangle (2,4.75);
\node at (0,3.125) {\includegraphics[scale=0.25]{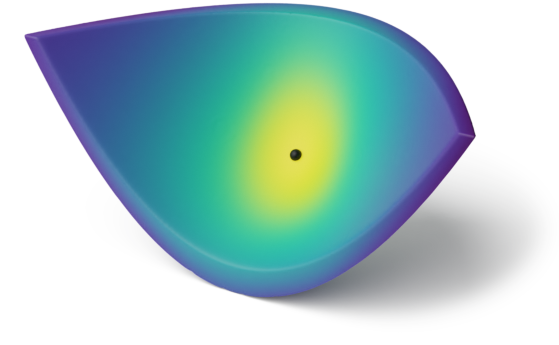}};
\node at (0,0) {\includegraphics[scale=0.25]{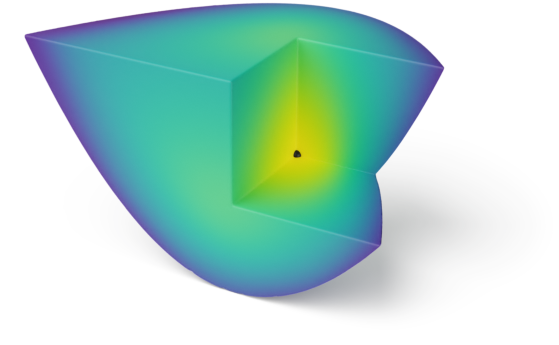}};
\end{tikzpicture}
\caption{$\kappa = 0.25$}
\end{subfigure}
\begin{subfigure}{0.3\textwidth}
\begin{tikzpicture}
\path[use as bounding box] (-2.375,-1.5) rectangle (2,4.75);
\node at (0,3.125) {\includegraphics[scale=0.25]{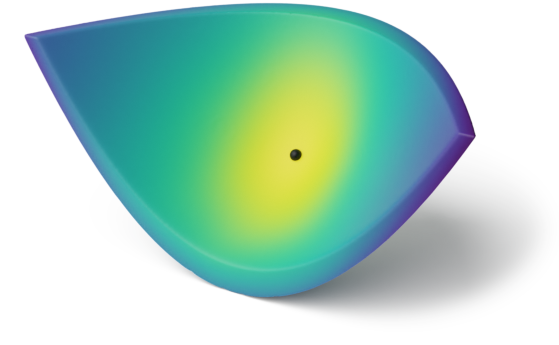}};
\node at (0,0) {\includegraphics[scale=0.25]{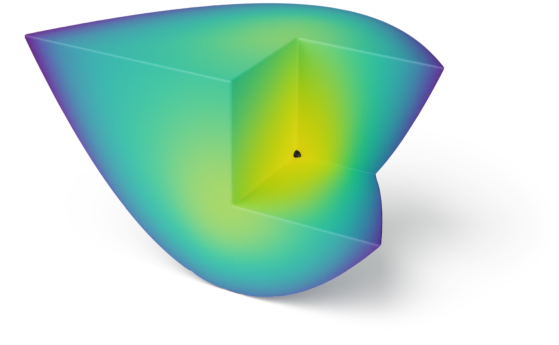}};
\end{tikzpicture}
\caption{$\kappa = 0.5$}
\end{subfigure}
\caption{Values of Matérn-5/2 kernels $k(\bullet,\.)$ on $\SPD(2)$, with various length scales.}
\label{fig:matern-values}
\end{figure}
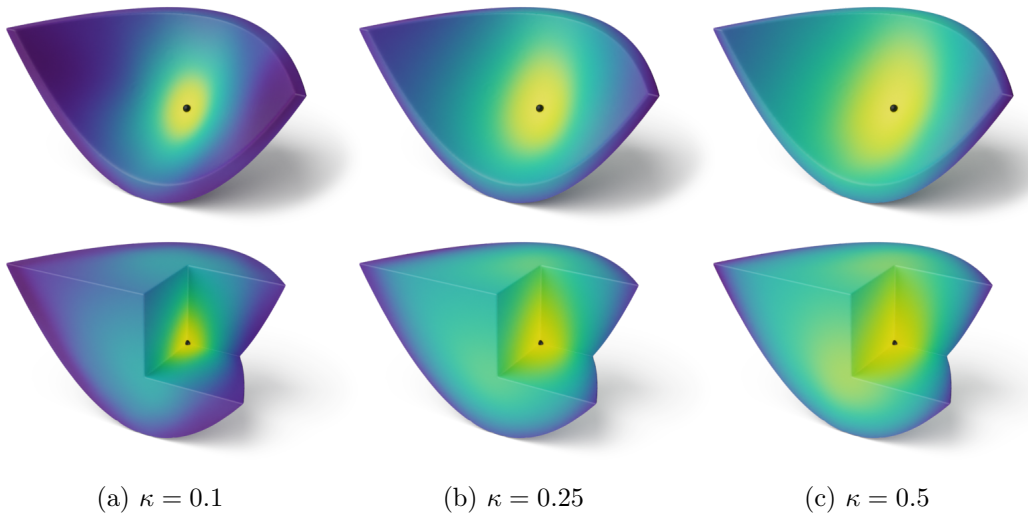

The smooth manifold $\SPD(d)$ has the same dimension as the space of symmetric matrices.
In the case $d=2$, this is a three-dimensional manifold, which can be understood as the set of all matrices $\begin{psmallmatrix}a & b \\ b & c\end{psmallmatrix}$ satisfying $a > 0$ and $ac-b^2>0$.
Geometrically, this space is a cone: to illustrate how Matérn kernels reflect the geometry of $\SPD(d)$, we visualize them for various length scales in \Cref{fig:matern-values}. 
We show cross-sections that correspond, respectively, to unit-determinant matrices, and central slices that correspond to entries equal to $a = 1$, $c = 1$, and $b = 0$.
The kernel's first argument is fixed at the identity matrix.

The geodesic distance induced by the affine-invariant metric \cite[Theorem~6.1.6]{bhatia2009} is given by
\[ \label{eqn:affine_inv_dist}
\f{dist}_{\SPD(d)}(\m{S}_1, \m{S}_2)
=
\norm{\log\del{\m{S}_1^{-1/2} \m{S}_2 \m{S}_1^{-1/2}}}_F
\]
where $\norm{\.}_F$ is the Frobenius norm of a matrix.
This distance is invariant under linear transformations, namely
\[
\f{dist}_{\SPD(d)}(\m{A} \m{S}_1 \m{A}^\top, \m{A} \m{S}_2 \m{A}^\top) = \f{dist}_{\SPD(d)}(\m{S}_1, \m{S}_2)
.
\]
It can therefore be extended to the space of Gaussian measures: in doing so, the above invariance transforms into an invariance with respect to affine transformations---this property gives it the name \emph{affine-invariant}.

\subsubsection{The Geometry of \texorpdfstring{$\SPD(d)$}{SPD(d)}} \label{sec:spd_geometry}

Consider first the case $d=1$: in this case, symmetric positive definite matrices are merely positive reals $\R_{>0}$, which form an abelian Lie group under multiplication.
Every abelian Lie group is isomorphic to a product of an Euclidean addition group with a torus: in this particular case, $\SPD(1)$ is isomorphic to $\R$ via the logarithm map $\log: \R_{>0} \-> \R$, which preserves the group structure.
This allows us to identify kernels on $\SPD(1)$ as kernels on $\R$ composed with the logarithm.
From now on we study the higher-dimensional case, where the group $\R_{>0}$ will play an important role: assume henceforth that $d > 1$.

We now proceed to develop a more precise understanding of the geometric structure of $\SPD(d)$, including how to give it the structure of a symmetric space, in order to apply the preceding ideas.
The following proposition shows how one can view $\SPD(d)$ as a homogeneous space, which in this form is \emph{almost}, but not quite, symmetric.

\begin{restatable}{proposition}{SPDHomogeneous}
\label{thm:spd_homogeneous}
The group $G = \GL(d)$ acts on $\SPD(d)$ transitively by conjugation as
\[
\label{eqn:spd_action}
\lacts : \GL(d) \x \SPD(d) &\-> \SPD(d)
&
\m{M} \lacts \m{S} &= \m{M}\m{S}\m{M}^\top
\]
where $H = \Ort(d) \subseteq G$ is the isotropy subgroup corresponding to the element $\m{I}_d \in \SPD(d)$.
$\SPD(d)$ can thus be regarded as the homogeneous space $\GL(d) / \Ort(d)$.
Moreover, the map
\[
\label{eqn:spd_s}
s:\GL(d) &\-> \GL(d)
&
s(\m{M}) &= \m{M}^{-\top}
\]
is an involution, namely $s(s(\m{M})) = \m{M}$, and $s(\m{M}) = \m{M}$ if and only if $M \in \Ort(d)$.
Using this representation, $\SPD(d)$ satisfies all aspects of the definition of a symmetric space, except for connectedness of the group $G$.
\end{restatable}

\begin{proof}
This is a standard result: we present its proof in~\Cref{apdx:proofs}.
\end{proof}

Unfortunately, the resulting quotient $\SPD(d) \isom \GL(d) / \Ort(d)$ is not a symmetric space: the first problem is that $\GL(d)$ is not connected.
This is relatively easy to fix, by restricting from $\GL(d)$ to its identity-connected-component subgroup $\GL_+(d)$, and, using the same argument as above, representing $\SPD(d) = \GL_+(d)/\SO(d)$.
This shows that $\SPD(d)$ is actually a symmetric space, but is still not good enough for our purposes since this representation fails to satisfy \Cref{asm:symmetric_space_assumptions}, as $\GL_+(d)$ is not semi-simple. 
To see this, note that a semi-simple Lie group cannot contain a normal abelian subgroup, like the subgroup of $\GL_+(d)$ formed by the matrices $a \m{I}_d$, $a > 0$.

To sidestep these issues, the idea will be to follow the strategy previously mentioned in the context of \Cref{asm:symmetric_space_assumptions}, and view $\SPD(d)$ as the product of the space of unit-determinant positive definite matrices $\SSPD(d) \subseteq \SPD(d)$ with the multiplicative group of positive reals $\R_{>0}$, representing the determinant, which already appeared when considering $\SPD(1)$.
To proceed, we show how $\SSPD(d)$ can be regarded as a symmetric~space.

\begin{restatable}{proposition}{SSPDSymmetric}
We have $\SSPD(d) \isom \SL(d)/\SO(d)$. 
Moreover, this space is symmetric and satisfies~\Cref{asm:symmetric_space_assumptions}.
\end{restatable}

\begin{proof}
This is a standard result: we present its proof in~\Cref{apdx:proofs}.
\end{proof}

Since $\SSPD(d)$ satisfies~\Cref{asm:symmetric_space_assumptions}, it possesses a canonical Riemannian metric induced by its Killing form.
The set $\R_{>0}$ of positive reals also possesses a canonical metric arising from pulling back the Euclidean metric on $\R$ through the logarithm map.
Thus, we can identify $\SPD(d)$ with the product of two Riemannian manifolds $\R_{>0}\x \SSPD(d)$, which we equip with the product metric to define a metric on $\SPD(d)$.
This metric is precisely the \emph{affine-invariant} metric we began with, which therefore arises from the symmetric space~structure.

\begin{restatable}{proposition}{SPDAsProduct}
Let $\R_{>0}$ and $\SSPD(d)$ be equipped with their respective canonical metrics, and let $\SPD(d)$ be equipped with the affine-invariant metric.
Then
\[
\c{T}: \R_{>0} \x \SSPD(d) &\-> \SPD(d)
&
\c{T} (a, \m{S}) &= a^{\sqrt{d}} \, \m{S}
\]
is a Riemannian isometry.
\end{restatable}

\begin{proof}
This is a known result, but we present its proof in~\Cref{apdx:proofs}.
\end{proof}

\subsubsection{Explicit Formulas for Spherical Functions and Spectral Measures}

We have established that $\SPD(d) = \R_{>0} \x (\SL(d) / \SO(d))$ is the product of the abelian group $\R_{>0}$ with a symmetric space.
To define the kernels of interest, formally speaking, we now need to (1) apply \Cref{sec:computation} together with standard Euclidean theory to define and compute the heat kernels on $\SSPD(d)$ and $\R_{>0}$, then (2) use the fact that the heat kernel on a product of Riemannian manifolds is the product of the heat kernels corresponding to its components \cite[Theorem~9.11]{grigoryan2009}, and finally (3) extend this to Matérn kernels by leveraging their connection to the heat kernel.

\textcite[Section 1.3]{terras2016} shows that the above calculation produces the same results that one would obtain if one were to simply \emph{pretend} that $\GL(d)$ is semi-simple, consider $\SPD(d) = \GL(d) / \Ort(d)$, and use the RQ decomposition\footnote{The RQ decomposition is defined similarly to the QR decomposition, but the two are distinct.} in lieu of the Iwasawa decomposition.
We will present the obtained results in these terms.
The vector $\rho$ from~\Cref{thm:harish-chandra} is given by \textcite[Section 1.3.1]{terras2016}\footnote{Note that in \textcite{terras2016} $\rho$ is two times smaller, this is because $\v{\lambda} = 2i\. \v{r}$, where vector $\v{r}$ in the book is called \emph{properly normalized variables in the power function.}} as
\[
\v{\rho}
=
\del{1-\frac{d+1}{2},\ldots,j-\frac{d+1}{2},\ldots,d-\frac{d+1}{2}}
.
\]
The zonal spherical functions are
\[
\pi^{\v{\lambda}}(\m{M}) &= \int_{\Ort(d)} p_{\v{\lambda}}(\m{H} \bdot \m{M})\d\mu_{\Ort(d)}(\m{H})
&
\m{M} &\in \GL(d)
\]
where, denoting by $u_j$ the $j$-th diagonal entry of the R part of the RQ decomposition of a matrix $\m{U} \in \GL(d)$, we have
\[
p_{\v{\lambda}}(\m{U})
=
\prod_{j=1}^d e^{(i\. \lambda_j+j-\frac{d+1}{2})\log u_{j}} = \prod_{j=1}^d u_{j}^{i\. \lambda_j+j-\frac{d+1}{2}}.
\]
Finally, Harish-Chandra's $c$-function is given by
\[
\label{eqn: psd cfunction}
|c(\v{\lambda})|^{-2} \propto \prod_{1\le i<j\le d} \pi|
\lambda_i-\lambda_j|\tanh(\pi|\lambda_i-\lambda_j|).
\]

\subsubsection{Sampling From Spectral Measures}
\label{sec:spd_evaluation}

We now turn attention to sampling from the spectral measure, which is the final missing piece on $\SPD(d)$.
Here, again, we concentrate on the ordinary Laplace--Beltrami operator, commenting on the changes required for the shifted Laplacian at the end of the section.
The respective spectral measures of heat and Matérn kernels on $\SPD(d)$ are given by
\[
\mu_{\nu,\kappa,\sigma^2}
\propto
\begin{cases}
\exp\del{-\frac{\kappa^2}{2}(\norm{\v{\lambda}}^2+\norm{\v{\rho}}^2)}
\quad
\prod\limits_{\mathclap{1\le i<j\le d}\vphantom{\strut}}
\quad
\pi |\lambda_i-\lambda_j|
\tanh(\pi|\lambda_i-\lambda_j|)
\d \v{\lambda}
\quad
&
\nu = \infty
\\
\del{\frac{2 \nu}{\kappa^2} + \norm{\v{\lambda}}^2 + \norm{\v{\rho}}^2}^{-\nu-\mathrlap{n/2}}
\quad
\prod\limits_{\mathclap{1\le i<j\le d}}
\quad
\pi| \lambda_i-\lambda_j|
\tanh(\pi|\lambda_i-\lambda_j|)
\d \v{\lambda}
\quad
&
\nu < \infty
\end{cases}
\]
where $\norm{\v{\rho}}^2 = \frac{d^3-d}{12}$ and $n = d(d+1)/2$.
We now observe these expressions are very closely related to the eigenvalue distribution of the following classical random matrix ensemble.

\begin{lemma}
\label{thm:GOE}
Define the \emph{Gaussian orthogonal ensemble} to be $\m{M} = (\m{X} + \m{X}^\top)/2$, where $\m{X}$ is a matrix with independent standard Gaussian entries.
Let $\v{\lambda}_{\f{GOE}}$ be a random vector consisting of eigenvalues of $\m{M}$ and denote its density by $p_{\v{\lambda}_{\f{GOE}}}$.
Then
\[
p_{\v{\lambda}_{\f{GOE}}}(\v{\lambda})
\propto
\exp\del{-\frac{\norm{\v{\lambda}}^2}{2}}
\prod_{1\le i<j\le d}
|\lambda_i-\lambda_j|
\]
\end{lemma}

\begin{proof}
\textcite[Theorem 3.3.1]{metha}.\footnote{We use a slightly different normalization compared to \textcite{metha}.}
\end{proof}

An immediate corollary of~\Cref{thm:GOE} is the following.

\begin{corollary} \label{thm:spd_heat_spectral}
Consider the density defined by
\[
p_{\infty, \kappa}(\v{\lambda})
&\propto
\exp\del{-\frac{\kappa^2}{2}(\norm{\v{\lambda}}^2+\norm{\v{\rho}}^2)}
\prod\limits_{1\le i<j\le d}
\pi |\lambda_i-\lambda_j|
&
\v{\lambda} &\in \R^d
.
\]
If $\v\lambda^\infty \~ p_{\v{\lambda}_{\f{GOE}}}$ and $\v\lambda^{\infty,\kappa} = \v\lambda^\infty/\kappa$, then $\v\lambda^{\infty,\kappa} \sim p_{\infty, \kappa}$. 
\end{corollary}
\begin{proof}
Let $p_{\v\lambda^{\infty,\kappa}}$ be the density of $\v\lambda^{\infty,\kappa}$.
Using the change of variable formula for densities
\[
p_{\v\lambda^{\infty,\kappa}}(\v\ell)
=
p_{\v\lambda^\infty}(\kappa\v\ell)\kappa^d.
\]
Hence $p_{\v\lambda^{\infty,\kappa}}(\v\ell) \propto p_{\infty, \kappa}(\v\ell)$, and since both are densities, they are equal.
\end{proof}

This is \emph{almost} the density which occurs in the heat kernel's spectral measure: once the length scale $\kappa$ is standardized and normalizing constants are dropped: all that remains is the hyperbolic tangent term.
Since $\tanh(|x|) < 1$, this connection can be used to devise an efficient rejection sampler, which behaves very similarly to the one previously developed for hyperbolic space.
As before, we first extend this to the Matérn class.

\begin{restatable}{proposition}{SPDSampling} \label{thm:spd_sampling}
Consider the density defined by
\[
p_{\nu, \kappa}(\v{\lambda})
&\propto
\del{
\frac{2 \nu}{\kappa^2}
+
\norm{\v{\lambda}}^2
+
\norm{\v{\rho}}^2
}^{-\nu-n/2}
\prod_{1\le i<j\le d}
\pi
|\lambda_i-\lambda_j|
&
\v{\lambda} &\in \R^d
.
\]
Let $\v\lambda^{\infty,\gamma} \sim p_{\infty,\gamma}$, where $\gamma = \del{\frac{2 \nu}{\kappa^2} + \norm{\v{\rho}}^2}^{-1/2}$ and the right-hand side is defined in~\Cref{thm:spd_heat_spectral}, and let $y \~ \chi_{2 \nu}$ be a chi-distributed random variable with $2 \nu \in \R_{>0}$ degrees of freedom.
If $\v\lambda^{\nu,\kappa} = \v\lambda^{\infty,\gamma}/y$, then $\v\lambda^{\nu,\kappa} \~ p_{\nu, \kappa}$.
\end{restatable}

\begin{proof}
Direct computation, see \Cref{apdx:proofs}.
\end{proof}

Using this, we obtain an analogous rejection-sampling-based strategy for sampling from the Matérn spectral measure.

\begin{algorithm}
Rejection sampler for $\mu_{\nu,\kappa,\sigma^2}$ on $\SPD(d)$.
\1  Sample a matrix $\m{X}$ whose entries are IID $\c{N}(0,1)$ and compute $\m{M}=(\m{X}+\m{X}^\top)/\sqrt{2}$.
\2 Compute the eigenvalues $\v{\lambda}^\infty$ of $\m{M}$.
\3 This step depends on whether or not $\nu$ is equal to $\infty$.
\1 If $\nu = \infty$, compute $\v{\lambda}^{\infty,\kappa} = \v{\lambda}^\infty/\kappa$.
\2 If $\nu < \infty$, compute $\v{\lambda}^{\infty,\gamma} = \v{\lambda}^\infty/\gamma$ where $\gamma = \del{\frac{2 \nu}{\kappa^2} + \norm{\v{\rho}}^2}^{-1/2}$ and sample $y \sim \chi_{2 \nu}$ from a chi distribution with $2 \nu$ degrees of freedom, compute $\v{\lambda}^{\nu, \kappa}=\v{\lambda}^{\infty, \gamma}/y$.
\0
\4 Reject $\v{\lambda}^{\nu,\kappa}$ with probability equal to $1-\prod_{1\le i<j\le d} \tanh \pi|\lambda_i-\lambda_j|$ returning to step 1, otherwise output $\v{\lambda}^{\nu,\kappa}$.
\0
\end{algorithm}

\begin{remark}
Setting $\gamma = \del{\frac{2 \nu}{\kappa^2}}^{-1/2}$ instead of $\gamma = \del{\frac{2 \nu}{\kappa^2} + \norm{\v{\rho}}^2}^{-1/2}$ results in an algorithm for sampling the measure $\h{\mu}_{\nu, \kappa, \sigma^2}$ corresponding to the shifted Laplacian, instead of $\mu_{\nu, \kappa, \sigma^2}$.
\end{remark}

This completes our development: on $\SPD(d)$, we can now compute heat and Matérn kernels, and perform efficient sampling from the prior.

\subsubsection{Alternative Expressions for Heat Kernels in Special Cases} \label{sec:spd_alt_expr}

For $d = 2$, there is a relatively simple alternative formula due to \textcite{sawyer1992}, which appeared in \textcite{jaquier2022}.
Take $\m{X}, \m{Y} \in \SPD(d)$ and denote by $\m{S}_\m{X}, \m{S}_\m{Y}$ their respective Cholesky factors.
Let $H_1 \geq H_2$ be the singular values of $\m{S}_\m{X} \m{S}_\m{Y}^{-1}$ and define $\alpha = H_1 - H_2$.
Then we have
\[ \label{eqn:spd_heat}
k_{\infty, \kappa, \sigma^2}(\m{X}, \m{Y})
=
\frac{\sigma^2}{C_{\infty}} \exp\del{-\frac{H_1^2 + H_2^2}{2 \kappa^2}} \int_0^{\infty} \frac{(2 s + \alpha) \exp\del{-s (s + \alpha)/ \kappa^2}}{\del{\sinh(s) \sinh(s + \alpha)}^{1/2}} \d s
\]
where $C_{\infty}$ is a normalizing constant ensuring that $k_{\infty, \kappa, \sigma^2}(\m{X}, \m{X}) = \sigma^2$.
Note that this expression, when implemented numerically using for instance quadrature, is not guaranteed to be positive semi-definite.

The expression \Cref{eqn:spd_heat} and the corresponding hyperbolic space expression \Cref{eqn:hyp_heat} look similar: this is not a coincidence.
Following \Cref{sec:spd_geometry}, we have $\SPD(2) = \R_{>0} \x \SSPD(2)$.
If $d=2$, and only in this case, then $\SSPD(d) = \bb{H}_d$ \cite[Introduction, \S4]{ggahelgason2000}, allowing one to apply techniques which are otherwise specific to hyperbolic spaces.
Leveraging this connection, one can obtain formulas like~\Cref{eqn:spd_heat}.
One can also derive expressions for $d=3$, which are unfortunately rather complicated: see \textcite{sawyer1992} for additional details.

\subsection{Approximation Error, Computational Costs, and Practical Considerations} \label{sec:experiments}

Since kernels on non-compact symmetric spaces admit integral representations, computing these kernels numerically generally involves Monte Carlo sampling, as opposed to truncation of infinite series which occurs in the compact case.
We now explore these Monte Carlo approximations and some practical considerations which arise from non-compactness.

\begin{figure}
\begin{subfigure}{0.33\textwidth}
\begin{tikzpicture}
\path[use as bounding box] (-2.625,-1.5) rectangle (2.125,1.5);
\node at (0,0) {\includegraphics[scale=0.25]{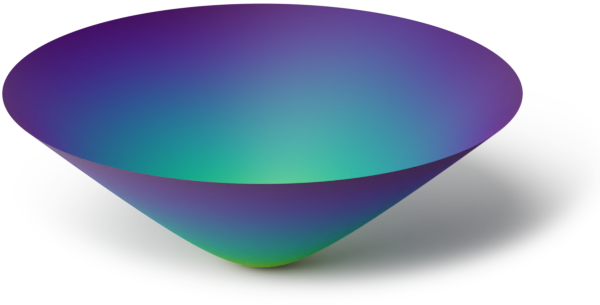}};
\end{tikzpicture}
\caption{Hyperbolic space}
\end{subfigure}
\begin{subfigure}{0.66\textwidth}
\begin{tikzpicture}
\path[use as bounding box] (-2.625,-1.5) rectangle (6.875,1.5);
\node at (0,0) {\includegraphics[scale=0.25]{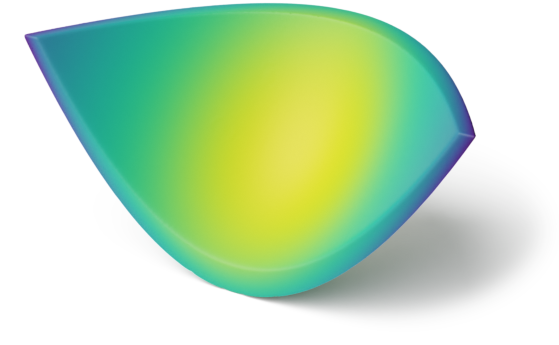}};
\node at (4.5,0) {\includegraphics[scale=0.25]{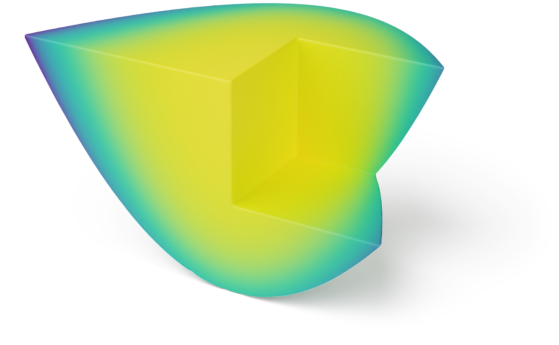}};
\end{tikzpicture}
\caption{Symmetric positive-definite matrices}
\end{subfigure}
\caption{The smallest zonal spherical function $\pi^{(0)}$, which can be non-constant.}
\label{fig:smallest-zonal-spherical-function}
\end{figure}

\subsubsection{Correlations in the Large Length Scale Limit}

The first important practical consideration which occurs for heat and Matérn kernels on non-compact symmetric spaces is that, in the large length scale limit, the correlation between any two given points need not converge to one.
This behavior, which differs from both the Euclidean and compact Riemannian settings, is a consequence of the following proposition.

\begin{restatable}{proposition}{PropKernelUpperBound}
Let $X = G/H$ be a symmetric space that satisfies~\Cref{asm:symmetric_space_assumptions}.
Recall that $\pi^{(\lambda)}$ denote zonal spherical functions on $X$ and $\lambda\in\fr{a}^*$ are vectors.
Consider $\Bbbk \in L^2(G)$ such that $k(g_1\bdot H, g_2\bdot H) = \Bbbk(g_2^{-1} \bdot g_1)$ is a stationary kernel on $X$.
Then
\[
\Bbbk(g)
\leq
\ubr{\Bbbk(e)}_{\t{variance of $k$}}
\pi^{(0)}(g)
\]
for all $g \in G$. 
Moreover, there exists $g \in G$ such that $g \not\in e \bdot H$ and $\pi^{(0)}(g) < 1$.
\end{restatable}

\begin{proof}
\Cref{apdx:proofs}.
\end{proof}

That is, under our usual assumptions, all stationary kernels are bounded above by the zonal spherical function $\pi^{(0)}$ which can be strictly less than one at points outside of the identity orbit~$e \bdot H$.
In the Euclidean or compact Riemannian cases, the analog of $\pi^{(0)}$ would be the constant function $1$, and this phenomenon does not occur.
\Cref{fig:smallest-zonal-spherical-function} shows $\pi^{(0)}$ for hyperbolic space and the space of symmetric positive-definite matrices.

\begin{figure}[t]
\includegraphics{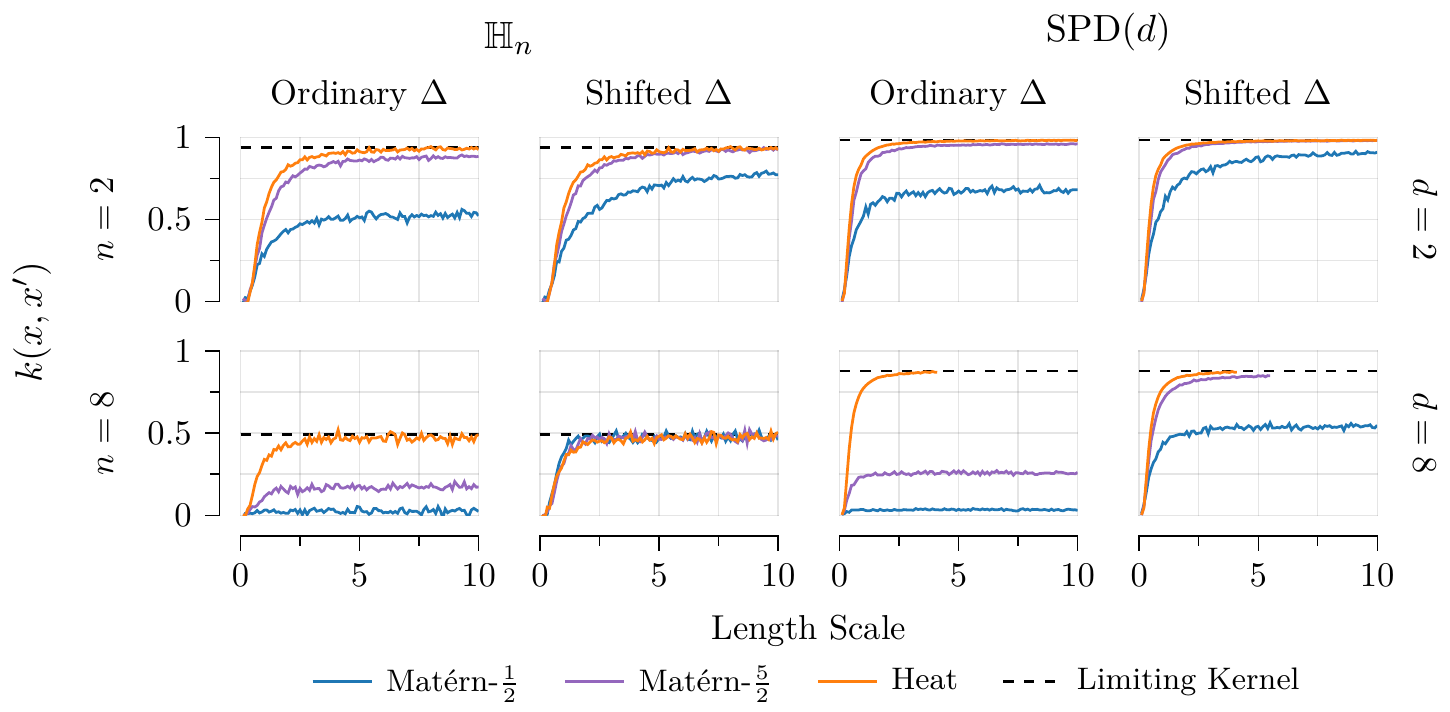}
\caption{Illustration of how $k(x,x')$ changes as $\kappa$ increases. To reduce noise stemming from approximation error, we average over $x, x'$ for which $k(x, x')$ is theoretically the same. We see that behavior depends on the smoothness and choice of Laplacian: for the ordinary Laplacian, or $\nu$ small enough, the limiting value may be significantly smaller than the value of the \emph{limiting kernel} $k^{(0)}(g_1 \protect\bdot H, g_2 \protect\bdot H) = \pi^{(0)}(g_2^{-1} \protect\bdot g_1)$. In contrast, kernels defined using the shifted Laplacian with $\nu$ large enough are guaranteed to reach this limit, albeit with a rate of convergence that depends on the geometry of the space and the kernel's smoothness, where smoother kernels tend to converge to their respective limiting values faster.}
\label{fig:range}
\end{figure}

Overall, the behavior of heat and Matérn kernels---whether one works with the shifted Laplacian, or not---is similar to that of Euclidean or compact Riemannian kernels, except for the aforementioned large length scale behavior.
For Matérn kernels specifically, this behavior also depends on the choice of Laplacian.
Recall that the density $p_{\nu,\kappa}(\lambda)$ of the measure $\mu_{\nu, \kappa, \sigma^2}$ with respect to the measure $|c(\lambda)|^{-2} \d\lambda$ is
\[
p_{\nu,\kappa}(\lambda)
\propto
\del{\frac{2 \nu}{\kappa^2} + \norm{\lambda}^2 + \norm{\rho}^2}^{-\nu-n/2}
\]
where $\nu < \infty$.
If we had $\norm{\rho}^2 = 0$ and $c(\lambda) = 1$ as in the Euclidean case, then $\mu_{\nu, \kappa, \sigma^2}$ would be the $t$-distribution, which converges to Dirac delta function as $\kappa\->\infty$.
The respective kernel, when evaluated at $g_1 \bdot H, g_2 \bdot H$, would then converge to $\pi^{(0)}(g_2^{-1}\bdot g_1)$---the closest thing one can get to a constant in our setting.
This is indeed what happens when the shifted Laplacian is used, granted that $\nu$ is large enough for the singularity of $p_{\nu,\kappa}$ to be stronger than the zero of $c(\lambda)^{-2}$.
However, this Euclidean-like behavior never happens when the ordinary Laplace--Beltrami operator is used.
This claim is formalized in the following proposition.

\begin{restatable}{proposition}{PropSpectralMeasureNonDirac} \label{thm:prop_spectral_measure_non_dirac}
For all $\nu<\infty$, the spectral measure $\mu_{\nu,\kappa, \sigma^2}$ does not converge weakly to a Dirac measure as $\kappa\->\infty$.
In contrast, if either $\nu = \infty$, or if we instead work with $\h{\mu}_{\nu,\kappa, \sigma^2}$ and $\nu$ is large enough, then the respective measure weakly converges to a Dirac measure.
\end{restatable}

\begin{proof}
\Cref{apdx:proofs}.
\end{proof}

\begin{figure}
\includegraphics{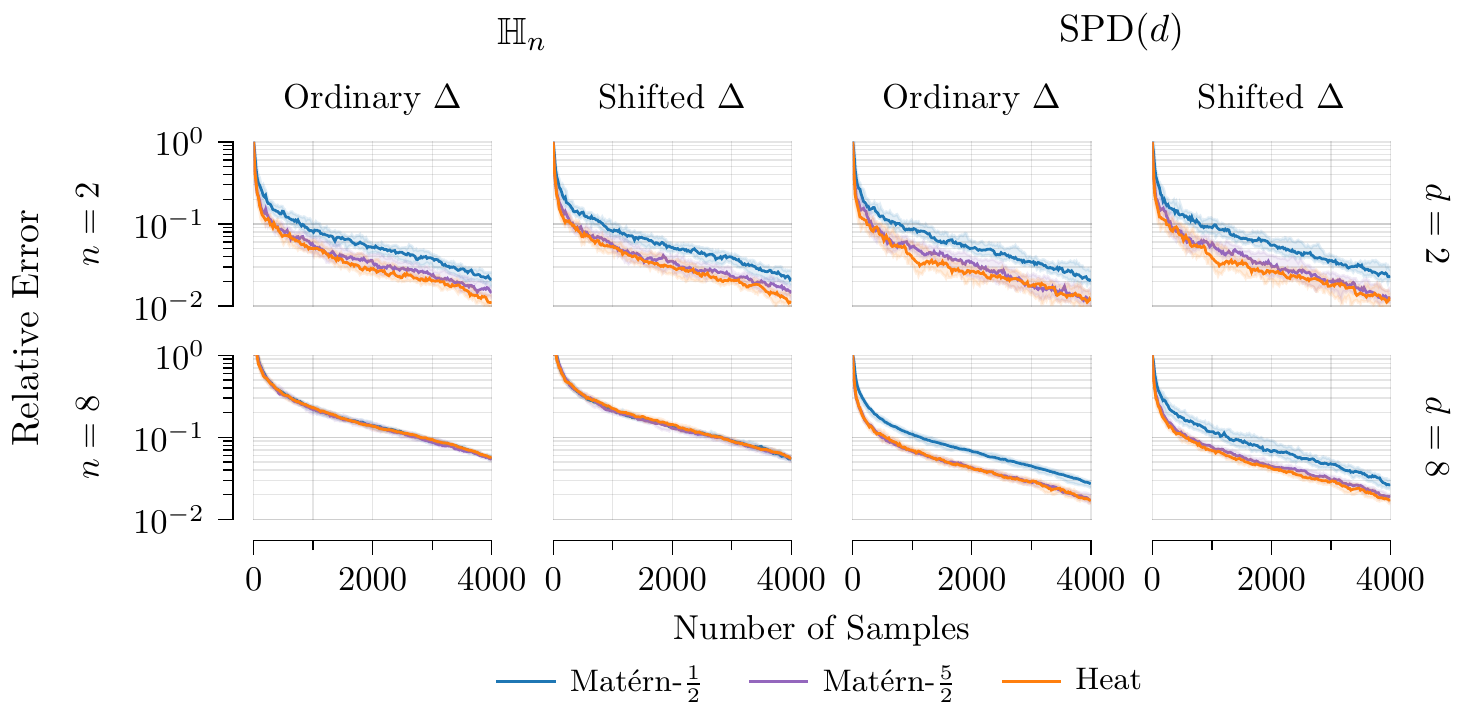}
\caption{Approximation error for the random Fourier feature approximation \Cref{eqn:approx_evaluation_pd_recall}, as a function of the number of Monte Carlo samples. Specifically, we plot relative error, which is equivalent to selecting a normalization for each curve so that its value at the origin equals one. We display the median, along with 25\% and 75\% quantiles, calculated via 20 different random seeds. We see that error decreases at the rate expected given the Monte Carlo nature of the approximation.}
\label{fig:rff_monte_carlo_error}
\end{figure}

The practical effect of this difference is that it further restricts the ability of Matérn kernels to capture large length scale behavior.
The reason for this difference is that, in the heat kernel, the term $\norm{\rho}^2$ enters the spectral density in a multiplicative manner that can be absorbed into the normalizing constant.
This does not occur in Matérn kernels defined using the ordinary Laplacian.
If one instead works with Matérn kernels defined using the shifted Laplacian, the term $\norm{\rho}^2$ vanishes, giving 
\[
\h{p}_{\nu,\kappa}(\lambda)
\propto
\del{\frac{2 \nu}{\kappa^2} + \norm{\lambda}^2}^{-\nu-n/2}
\]
which resembles the Euclidean case more closely.
This means that, Matérn kernels defined using the shifted Laplacian are better able to represent functions which are more spatially constant, and any remaining limitations in this ability arise from the geometry of the space, that is, from the zonal spherical functions $\pi^{(\lambda)}$, and not from the choice of kernel.
This provides a practical reason to work with the shifted rather than ordinary Laplacian.

We conclude by examining large length scale behavior empirically.
\Cref{fig:range} shows how $k(x,x')$, for a fixed set of points $x \neq x'$, changes as one varies the length scale, for the hyperbolic space and space of symmetric positive definite-matrices examples developed in this work.
We see that, if the ordinary Laplacian is used, $k(x,x')$ can converge to a significantly smaller value, compared to if the shifted Laplacian is used---especially for less-smooth kernels.
We also see that the rate of convergence depends on the interplay between the geometry of the space, the kernel's smoothness, and the choice of Laplacian.
We will examine this behavior further in the sequel, in the context of rejection sampler efficiency---to do so, we proceed to discuss further approximation considerations.
To conclude, note that $\bb{H}_2$ and $\SPD(2)$ behave similarly: this occurs because of the factorization discussed in \Cref{sec:spd_alt_expr} which connects the two spaces.
In contrast, $\bb{H}_8$ and $\SPD(8)$ exhibit a wider range of behavior.

\begin{figure}
\includegraphics{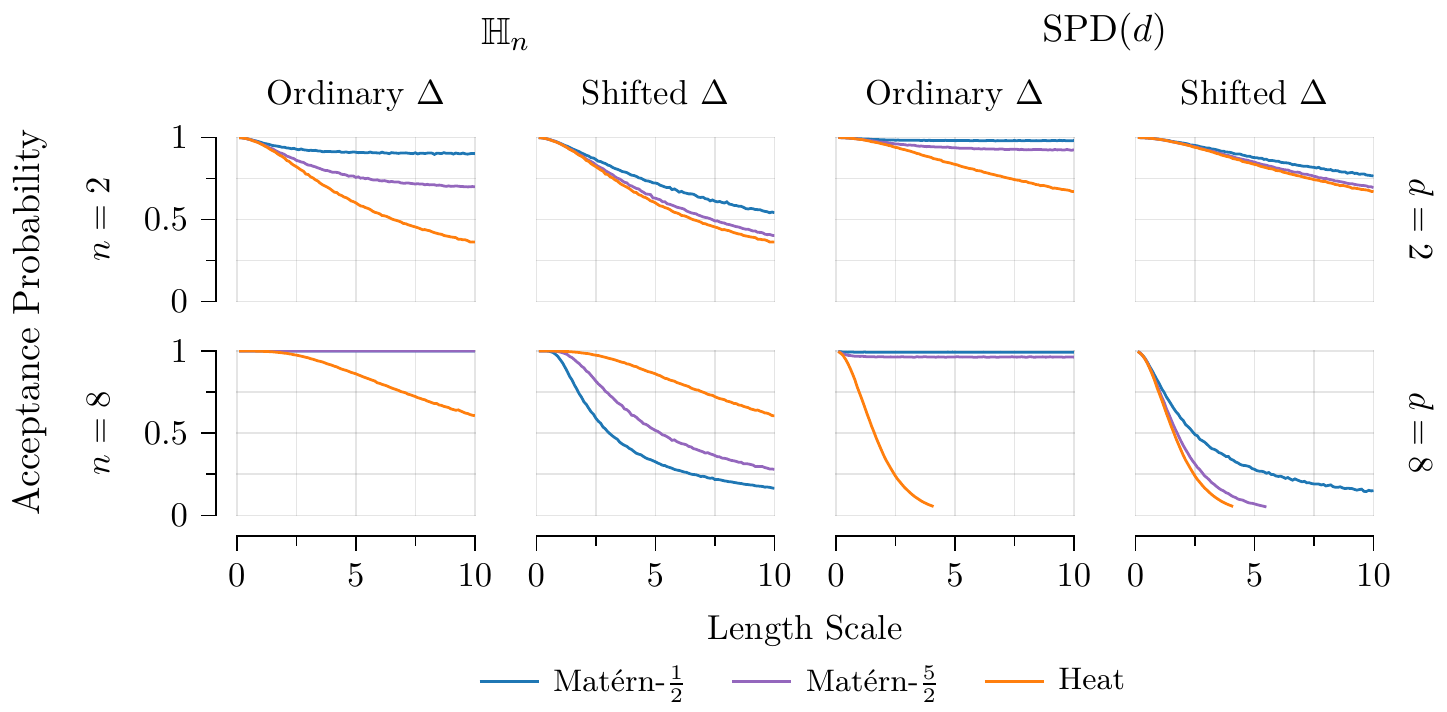}
\caption{Variability in acceptance probability of the rejection samplers for the respective Matérn and heat kernels' spectral measures, depending on smoothness, length scale, and dimension of the space. Acceptance probabilities concentrate around one for small length scales, and decrease for larger ones. This rate of decrease is faster in higher dimension.}
\label{fig:rejection_sampler}
\end{figure}

\subsubsection{Empirical Approximation Error and Rejection Sampler Efficiency}

We now examine approximation error for the kernels considered.
We focus on the random Fourier feature approximation \Cref{eqn:approx_evaluation_pd}, given by
\[ 
\label{eqn:approx_evaluation_pd_recall}
k(g_1 \bdot H, g_2 \bdot H) &\approx \frac{\sigma^2}{L} \sum_{l=1}^L e^{(i\lambda_l+\rho) a(h_l\bdot g_1)}
\overline{e^{(i\lambda_l+\rho)a(h_l\bdot g_2)}}
&
\lambda_l &\~ \frac{1}{\sigma^2} \mu_k
&
h_l &\~ \mu_H
\]
because, unlike the approximation in~\Cref{eqn:approx_evaluation}, it is guaranteed to be symmetric and positive semi-definite for all $L$, and is thus a more practical choice for many applications.
By \Cref{thm:rff_convergence}, the variance of this approximation is uniformly bounded on compact subsets.
To get an idea for how accurate this approximation is, we evaluate it empirically on a set of hyperbolic spaces $\bb{H}_n$, and spaces of symmetric positive definite matrices $\SPD(d)$.

\Cref{fig:rff_monte_carlo_error} shows that approximation error, quantified using relative $L^2$-error from an approximate ground-truth kernel computed using $5000$ Monte Carlo samples.
This quantity is seen to decrease at the usual square-root Monte Carlo rate.

For Matérn and heat kernels, the approximation \Cref{eqn:approx_evaluation_pd_recall} requires sampling from the respective spectral measure, which for the spaces of interest---except $\bb{H}_n$ with odd $n$---involves rejection sampling.
To understand the efficiency of these samplers, we evaluate their acceptance probabilities for $\bb{H}_n$ and $\SPD(d)$, and examine how this changes as a function of the dimension of the space, as well as smoothness and length scale parameters $\nu$ and~$\kappa$.

\Cref{fig:rejection_sampler} shows results.
For sufficiently-small length scales, the proposed rejection sampling scheme is very efficient, with near-unit acceptance probabilities.
As the length scale increases, acceptance probabilities drop, with higher-dimensional spaces exhibiting a faster drop than lower-dimensional spaces.
This rate is smoothness-dependent: for $\bb{H}_2$, $\SPD(2)$, and $\SPD(8)$, the rejection sampler for the less-smooth Matérn-1/2 kernel is more efficient compared to that of the smoother heat kernel, while for $\bb{H}_8$ the opposite holds.
This indicates that efficiency depends on the interplay between the kernel and the geometry of the space.

While this may at first seem like a problem, observe by comparing \Cref{fig:rejection_sampler,fig:range}, that regions where acceptance probabilities are low are also regions where the kernel has converged to its large-length-scale limit.
In those cases, one therefore can simply approximate $k$ by its limit, and avoid Monte Carlo error as an extra benefit.
We conclude, empirically, that the proposed rejection samplers are inefficient in the same situations where they are~unneeded.

\section{Conclusion}

In this paper, we developed computationally tractable Gaussian process models on a natural class of non-compact Riemannian manifolds where this is possible, namely symmetric spaces.
We derived specific kernel expressions and sampling algorithms for two concrete classes of spaces which are important in applications: hyperbolic spaces and the manifolds of symmetric positive definite matrices.
Our work brings together a number of results from probability and statistics, differential geometry, harmonic analysis, and representation theory.
In total, we obtain kernel matrix approximations which are guaranteed to be positive semi-definite, making them simple and reliable to use. 
We hope our techniques enable Gaussian processes to be applied in a wider set of potential applications with geometric structure.

\section*{Acknowledgements}

The authors are grateful to Prof. Mikhail Lifshits for providing useful feedback.
IA and AS were supported by RSF grant N\textsuperscript{\underline{o}}21-11-00047.
AT was supported by Cornell University, jointly via the Center for Data Science for Enterprise and Society, the College of Engineering, and the Ann S. Bowers College of Computing and Information Science.
VB was supported by an ETH Zürich Postdoctoral Fellowship.
We further acknowledge support from Huawei Research and Development.

\appendix

\section{Additional Details on Non-compact Symmetric Spaces} \label{apdx:noncompact}

Here, provide further details on the geometric and algebraic structure of symmetric spaces.
We mostly follow \textcite{helgason1962differential, helgason1979differential, helgason1994}.
We begin with the arguably more intuitive geometric picture.

\begin{definition}
Let $M$ be a connected manifold with Riemannian metric $g$. 
For a point $p\in M$ let $T_p M$ be a tangent space at $p$. 
Let $s_p$ be the \emph{geodesic symmetry} defined on the image $U$ of a sufficiently small ball $B(0_p,\varepsilon)\subset T_p M$ under the exponential map evaluated at $p$---or, more precisely, by
\[
s_p \exp(v) &= \exp(-v) 
&
v &\in B(0_p,\varepsilon)
.
\]
The space $M$ is called a \emph{locally symmetric space} if, for all points $p\in M$, the map $s_p$ is an isometry. 
It is called a (globally) \emph{symmetric space} if this isometry can be extended to an isometry of the whole space $M$.
\end{definition}

Roughly speaking, this means that a Riemannian manifold $(M,g)$ is a symmetric space if all point reflections are isometries of the manifold. 
This holds for such spaces as spheres, Grassmannians, Euclidean, hyperbolic spaces, and many others.

From the definition of a symmetric space, it immediately follows that $M$ is geodesically complete, since every geodesic can be continued by reflection about its endpoints.
Geodesic completeness implies, by the Hopf--Rinow Theorem, that for every $p,q \in M$, we can find a geodesic $\gamma:[0,1]\-> M$ that connects $p$ and $q$, namely $\gamma(0)=p$ and $\gamma(1)=q$ \cite[Theorem~10.4]{helgason1979differential}.
Reflecting points around the midpoint of a geodesic, $\gamma(1/2)$, sends $p$ to $q$ and $q$ to $p$.
Thus, the group of isometries acts transitively on $M$.

Let $G$ be the isometry group of $M$.
By the Myers--Steenrod Theorem \cite{myers1939} $G$ can naturally be given the structure of a Lie group.
Fix an arbitrary point $p\in M$ on the manifold. 
One can show that $p$ generates a Lie subgroup $H$ of isometries that leaves $p$ fixed in the sense that $g \lacts p = g$ if and only if $g \in H$: call $H$ the \emph{isotropy group}. 
This allows one to identify $M$ as a homogeneous space $G/H$.

We can use this representation to study symmetric spaces in an algebraic, Lie-theoretic way.
The reflection map $s_p$ defines an involution $\sigma$ on $G$, namely
\[
\sigma(g) &= s_p \circ g \circ s_p
&
g &\in G
.
\]

The main technical tool for studying Lie groups is the notion of a \emph{Lie algebra}, which allows one to linearize various questions that concern Lie groups. 

\begin{definition}
A \emph{Lie algebra} $\fr{g}$ is a vector space equipped with a bilinear operation $[\.,\.]: \fr{g} \x \fr{g} \-> \fr{g}$, called the \emph{Lie bracket}, which is a bilinear form that is \emph{alternating}, namely $[X,X] = 0$, and satisfies the \emph{Jacobi identity}
\[
[X,[Y,Z]] + [Y,[Z,X]] + [Z,[X,Y]] = 0
.
\]
\end{definition}

Every Lie group gives rise to a Lie algebra, which is constructed as follows.

\begin{definition}
Let $G$ be a Lie group.
Define the \emph{Lie algebra of a Lie group} to be $\fr{g} = T_e G$, the tangent space of $G$ at the identity element $e$, equipped with the Lie bracket constructed as follows.
One can show that each element of $\fr{g}$ is in bijective correspondence with left-invariant vector fields on $G$, namely vector fields $X$ such that $(\D L_g)X_h = X_{g \bdot h}$, where $L_g: G\-> G$ is the left multiplication map  $L_g(h)=g \bdot h$. 
Every vector field $X$ defines a derivation, namely $X(Y)(g)$, which is a directional derivative of $Y$ along the tangent vector $X_g$.
With this notation, the Lie bracket of a pair of vector fields is defined as $[X,Y] = X(Y)-Y(X)$, which carries over to the Lie algebra by the aforementioned bijective correspondence.
\end{definition}

In the most important case of a linear Lie group, that is, a closed subgroup of $\GL(n)$, the tangent space is naturally identified with a subspace of the vector space of all matrices~$\f{M}(n,\mathbb{R})$.
Here, the Lie bracket is simply the additive commutator $[\m{X},\m{Y}]=\m{X}\m{Y}-\m{Y}\m{X}$.

Let $\fr{g}$ and $\fr{h}$ be the Lie algebras of $G$ and $H$, respectively, where $G$ and $H$ are as above.
Since $\sigma$ is an involution, meaning $\sigma^2$ is the identity map, the differential of $\sigma$ at the identity, $\theta = \D_e\sigma$ is also an involution, so it is a linear map with eigenvalues $\{+1,-1\}$.
The eigenspace corresponding to $+1$ is exactly $\fr{h}$. 
Let $\fr{p}$ be the eigenspace corresponding to $-1$, which can be identified with the tangent space $T_p M$ \cite[Theorem~3.3(iii)]{helgason1979differential}.
We call the pair $(\fr{h},\fr{p})$ a \emph{Cartan pair}.

Recall that a Lie algebra is called \emph{semi-simple} if it contains no non-zero abelian ideals.\footnote{A Lie algebra $\fr{g}$ is called \emph{abelian} if $[X,Y]=0$ for all $Y$ and $Y$ in $\fr{g}$. A subspace $\fr{i}$ of a Lie algebra $\fr{g}$ is called an \emph{ideal} if it is stable under the Lie bracket action of $\fr{g}$: that is, for any $X\in\fr{g}$ and $Y\in\fr{i}$ one has $[X,Y]\in\fr{i}$. This means that $\fr{i}$ is a subalgebra.}
Equivalently, a Lie algebra is semi-simple if it can be decomposed into a direct sum of \emph{simple}\footnote{A non-abelian Lie algebra is called \emph{simple} if it does not contain any non-trivial ideals.} Lie algebras. 
A Lie group is called \emph{semi-simple} if its Lie algebra is \emph{semi-simple}.
The principal source of examples of semi-simple Lie groups are the simple Lie groups, which are defined as connected non-abelian groups containing no connected normal subgroups---note that such groups can have a discrete normal subgroup, as $\SL(n)$ does, for example.

Our spaces of interest have the form $G/H$ where $G$ is a non-compact \emph{semi-simple} Lie group with \emph{finite center} and $H$ is a maximal, by inclusion, compact subgroup of $H$.
Most semi-simple Lie groups which occur in practice satisfy the latter assumption.
Let $\fr{g}$ and $\fr{h}$ be Lie algebras corresponding to $G$ and to $H$. 
Recall that we use $e$ to refer to the identity element of~$G$.
Define the \emph{adjoint action} of $X\in \fr{g}$ by
\[
\ad(X) : \fr{g} &\-> \fr{g},
&
\ad(X)(Y) &= [X,Y]
.
\]
where $[\.,\.]$ denotes the Lie bracket of $\fr{g}$. 
Define the \emph{Killing form} $B$ of $\fr{g}$ to be a symmetric bilinear form on $\fr{g}\x \fr{g}$ defined by
\[
B(X,Y) = \tr(\ad(X) \o \ad(Y))
\]
where $\o$ denotes composition of functions.
One can show that a non-zero Lie algebra is semi-simple if and only if its Killing form is non-degenerate.

The subalgebra $\fr{p}$ defined previously can be viewed as the orthogonal complement of $\fr{h}$ with respect to $B$.
The differential at $e$ of involution $\sigma$, for $X\in \fr{h}$ and $Y\in \fr{p}$, can be written
\[
\d_e\sigma(X+Y) = X-Y
.
\]

Within the subalgebra $\fr{p}$, let $\fr{a}$ be a maximal abelian subalgebra, and let $\fr{a}^*$ be its dual vector space.
For a linear functional $\alpha \in \fr{a}^*$ we can define the \emph{root space}
\[ \label{eqn:root_space}
\fr{g}_\alpha = \{X\in \fr{g} : [X,Y] = \alpha(Y)X,\, \forall Y\in \fr{a}\}
.
\]
This is a vector space: let $m_\alpha$ be its dimension.
If $m_\alpha > 0$ and $\alpha \neq 0$, then $\alpha$ is called a \emph{restricted root}.
Let $\Sigma$ be the set of restricted roots. 
One can show that the number of restricted roots is finite.
A detailed description of restricted roots and their multiplicities for irreducible symmetric spaces can be found in~\textcite{araki1962}.

From the restricted roots, one can choose a set of \emph{positive roots} $\Sigma^+$ defined by the following conditions.

\1 For each restricted root $\alpha$, exactly one of $\alpha$ and $-\alpha$ is contained in $\Sigma^+$.
\2 For any $\alpha,\beta\in \Sigma^+$, either $\alpha+\beta\in \Sigma^+$ or $\alpha+\beta$ is not a restricted root.
\0

With these notions, we can define Lie subalgebra $\fr{n}$ of $\fr{g}$ by
\[
\fr{n} = \bigoplus_{\alpha\in \Sigma^+} \fr{g}_\alpha
.
\]
Putting these components together, we arrive at the \emph{Iwasawa decomposition} of the Lie algebra $\fr{g}$, which is
\[
\fr{g} = \fr{n} \oplus \fr{a} \oplus \fr{h}
.
\]
This decomposition can be lifted from the Lie algebra to the Lie group.

\begin{theorem} 
Let $G$ and $H$ be defined as above and let $\fr{g} = \fr{n} \oplus \fr{a} \oplus \fr{h}$ be an Iwasawa decomposition of the Lie algebra $\fr{g}$ of $G$.
Then there are subgroups $A$ and $N$ of $G$, whose respective Lie algebras are $\fr{a}$ and $\fr{n}$, such that the multiplication map $N\x A\x H\-> G$ given by $(n,a,h)\mapsto n\bdot a\bdot h$ is diffeomorphism.
Moreover, the subgroups $A$ and $N$ are simply connected. 
\end{theorem}

For $g \in G$, let $N(g) \in N$, $A(g) \in A$, and $H(g) \in H$ be the elements for which $g = N(g) \bdot a(g) \bdot H(g)$.
We now illustrate a few examples.

\begin{example}
For $G=\SL(n)$ let $N$ be the subgroup of upper-triangular matrices with $1$ on the diagonal, $A$ be the subgroup of diagonal matrices with non-negative entries, and $H=\SO(n)$. Then the Iwasawa decomposition becomes the well-known RQ-decomposition.
\end{example}

\begin{example} \label{example:so1n_iwasawa}
Let $G=\SO_0(1, n)$ be the identity component of the Lorentz group---this group is introduced in~\Cref{sec:hyperbolic_space}. 
Its Lie algebra $\fr{g}$ takes the form
\[
\fr{g} & = \{ \m{X}\in \f{M}(n+1, \R) : \forall \v{u},\v{v} \in \R^{n+1}\ \innerprod{\m{X}\v{u}}{\v{v}}_M + \innerprod{\v{u}}{\m{X}\v{v}}_M = 0 \} \\
& = \{ \m{X}\in \f{M}(n+1, \R) : \m{X}^\top\m{B} + \m{B}\m{X} = 0 \} \\
& = \cbr{ \begin{pmatrix} 0 & \v{b}^\top \\ \v{b} & \m{D}\end{pmatrix} : \v{b}\in\R^n,\ \m{D}^\top=-\m{D}  }\label{eq:so1n_lie_algebra}
\]
where $\m{B} = \f{diag}(-1,1,\ldots,1)$ is the Gram matrix of $\innerprod{\.}{\.}_M$ defined in~\Cref{eqn:minkowski_product} and, as before, $\f{M}(n+1, \R)$ denotes the vector space of $(n+1) \x (n+1)$ matrices.

We now present a computational recipe for obtaining the Iwasawa decomposition, adapted from \textcite{sawyer2016}. 
Let $\m{Q}$ be the following orthogonal matrix:
\[
\m{Q} = \begin{pmatrix}
\frac{1}{\sqrt{2}} & \frac{1}{\sqrt{2}} & \v{0} \\
\frac{1}{\sqrt{2}} & \frac{-1}{\sqrt{2}} & \v{0} \\
\v{0} & \v{0} & \m{I}_{n-1}
\end{pmatrix}.
\]
If $\m{Q}^\top \m{M} \m{Q} = \m{N}\m{A}\m{H}$ is the Iwasawa decomposition inside $\SL(n+1)$, then the Iwasawa decomposition of $\m{M}\in G$ is
\[
\m{M} = \m{Q}\m{N}\m{Q}^\top \. \m{Q}\m{A}\m{Q}^\top \. \m{Q}\m{H}\m{Q}^\top = N(\m{M}) \. A(\m{M}) \. H(\m{M}).
\]
The resulting matrices can be described explicitly.
Firstly, $A(\m{M})$ has the following form, for some $t > 0$:
\[
\m{A}_t &= \exp(a(t)) =
 \begin{pmatrix}
\cosh{t} & \sinh{t} & \v{0} \\
\sinh{t} & \cosh{t} & \v{0} \\
\v{0} & \v{0} & \m{I}_{n-1}
\end{pmatrix}
&
a(t) &=
\begin{pmatrix}
0 & t & \v{0} \\
t & 0 & \v{0} \\
\v{0}& \v{0} & \m{0}_{n-1}
\end{pmatrix}
\in \fr{a}
\]
where $\m{0}_{n-1}$ is the $(n-1) \x (n-1)$ matrix of zeros.
For a matrix $\m{X}\in\fr{g}$ partitioned as
\[
\m{X} = \begin{pmatrix}
0 & a & \v{b}^\top \\
a & 0 & -\v{c}^\top \\
\v{b} & \v{c} & \m{D}
\end{pmatrix}
&&
\v{b}, \v{c} \in \R^{n-1}
&&
\m{D}^\top = -\m{D}
\]
the commutator $[\m{X},a(t)]$ takes the form
\[
[\m{X},a(t)] = t \.
\begin{pmatrix}
0 & 0 & \v{c}^\top \\
0 & 0 & -\v{b}^\top \\
\v{c} & \v{b} & \m{0}_{n-1}
\end{pmatrix}
.
\]
Equating this to $\alpha(a(t))\m{X}$ with $\alpha(a(t)) = \lambda t$ for some $\lambda$, one gets from the definition of the root spaces \Cref{eqn:root_space} that $\lambda=\pm1$. Choosing $\alpha(a(t))=t$ as the positive root results in the elements of $\fr{n}$ taking the form
\[
&
\begin{pmatrix}
0 & 0 & \v{b}^\top \\
0 & 0 & -\v{b}^\top \\
\v{b} & \v{b} & \m{0}_{n-1}
\end{pmatrix}
&
\v{b} &\in \mathbb{R}^{n-1}
\]
thus the matrix $N(\m{M})=\exp(\m{X})$ is of the form
\[
\m{N}_{\v{b}} &=
\begin{pmatrix}
1 + \Delta & \Delta & \v{b}^\top \\
-\Delta & 1-\Delta & -\v{b}^\top \\
\v{b} & \v{b} & \m{I}_{n-1}
\end{pmatrix}
&
\v{b} &\in\mathbb{R}^{n-1}
&
\Delta&=\frac{\norm{\v{b}}^2}{2}.
\]
\end{example}

The abelian Lie algebra $\fr{a}$ endowed with the inner product given by the restriction of the Killing form $B|_{\fr{a}\x \fr{a}}$ can be identified with the Euclidean space $\R^{l}$ for some $l$. We say that $\dim\fr{a} = l$ is the \emph{rank} of the symmetric space $G/H$.

A vector $X\in \fr{a}$ is called \emph{regular} if $\alpha(X)\neq 0$ for all $\alpha\in \Sigma$.
Since $\Sigma$ is a finite set, the connected components of the set of regular elements are convex cones, which are called \emph{Weyl chambers}.
By construction, the sign of $\alpha(X)$ is constant on each Weyl chamber. 
Let $\fr{a}_+$ be the \emph{positive Weyl chamber}, which is the unique Weyl chamber such that all elements of $\fr{a}_+$ take positive values on $\Sigma^+$. 
Define \emph{Weyl group} $W$ to be the finite subgroup of the group of bijective linear transforms $\GL(\fr{a})$, which is generated by reflections relative to the hyperplanes $L_\alpha =\{\alpha(X)=0| X\in \fr{a}\}$, where $\alpha\in \Sigma$. 
This group acts by permutations on the set of Weyl chambers.

Using the Killing form $B$, we can identify $\fr{a}$ and $\fr{a}^*$: for $\lambda\in \fr{a}^*$ we associate $X_\lambda\in \fr{a}$ such that $B(X_\lambda,X) = \lambda(X)$ for all $X\in \fr{a}$. 
Thus, the space $\fr{a}^*$ is naturally endowed with the inner product $\innerprod{\lambda}{\mu} = B(X_\lambda,X_\mu)$. 
Finally, the Weyl chamber and the Weyl group can lifted to $\fr{a}^*$ by writing
\[
\fr{a}^*_+ = \{\lambda\in \fr{a}^* : X_\lambda\in \fr{a}^+\}
.
\]
Similarly, the Weyl group $W$ is generated by reflections relative to the hyperplanes $L^*_\alpha = \{\alpha(X_\lambda)=0 : \lambda\in \fr{a}^*\}$, where $\alpha\in\Sigma$, which are defined on the dual space of $\fr{a}$.

\section{Spherical Fourier Transform} \label{apdx:spherical_fourier_transform}

Here we introduce the notion of Fourier transform appropriate for non-compact symmetric spaces.
We use notation and definitions from~\Cref{apdx:noncompact}.
Additionally, we define $L^p_H(G, \mu_G)$ to be the subspace of $L^p(G, \mu_G)$ which consists of $H$-bi-invariant functions, meaning all $f\in L^p(G,\mu_G)$ such that $f(h_1 \bdot g \bdot h_2) = f(g)$.

\begin{result}
\label{thm:fourier_transform}
The \emph{spherical Fourier transform} of $f\in L^1_H(G, \mu_G)$ is given by
\[ \label{eqn:spherical_fourier_transform}
\widehat{f}(\lambda) &= \int_G f(g)\overline{\pi^{(\lambda)}(g)} \d \mu_{G}(g)
\]
for $\lambda \in \fr{a}^*$, where the function $\widehat{f}$ is uniquely determined by its values on the positive Weyl chamber $\fr{a}^*_+ \subseteq \fr{a}^*$.
Moreover, every $f\in L^1_H(G, \mu_G)\cap L^2_H(G, \mu_G)$ can be recovered from its spherical Fourier transform using the inversion formula
\[
\label{sym: inv_formula}
f(g)
=
\int_{\fr{a}^*_+}\widehat{f}(\lambda)\pi^{(\lambda)}(g)|c(\lambda)|^{-2}\,\d\lambda
=
\frac{1}{\abs{W}}\int_{\fr{a}^*}\widehat{f}(\lambda)\pi^{(\lambda)}(g)|c(\lambda)|^{-2}\,\d\lambda
\]
where integration is taken with respect to the Lebesgue measure $\d \lambda$ on $\fr{a}^*$, $c(\lambda)$ is a space-dependent function called \emph{Harish-Chandra's $c$-function}, and $\abs{W}$ denotes the order of the Weyl group.
Finally, the map $f \|> \widehat{f}$ is an isometry between the spaces $L^2_H(G, \mu_G)$ and $L^2\del{\fr{a}^*_+,|c(\lambda)|^{-2}\,\d\lambda}$.
\end{result}

\begin{proof}
\textcite[Chapter III, Theorem 1.5]{helgason1994}.
\end{proof}

Crucially, the zonal spherical functions $\pi^{(\lambda)}$ are eigenfunctions of the Laplace--Beltrami operator on $G$~\cite{Gangolli1968}.
Thanks to this, and to the fact that the spherical Fourier transform is built from these zonal spherical functions, we get the following.

\begin{result}
\label{thm:heat_kernel_apdx}
The solution of the heat equation on a symmetric space $G/H$ is given by
\[ \label{eqn:heat_symmetric} 
\c{P}(t, g_1\bdot H, g_2\bdot H)
\propto
\int_{\fr{a}^*}
  e^{-t(\norm{\lambda}^2 + \norm{\rho}^2)}
  \.
  \pi^{(\lambda)}(g_2^{-1} g_1)
  \.
  |c(\lambda)|^{-2}
\d\lambda
\]
where $\rho = \frac{1}{2}\sum_{\alpha\in \Sigma^+} m_\alpha \alpha$ is the half-sum of all positive roots, integration is taken with respect to the Lebesgue measure on $\fr{a}^*$, and $c(\lambda)$ is Harish-Chandra's $c$-function.
\end{result}

\begin{proof}
The full proof can be found in \textcite{Gangolli1968}.
A sketch of the argument is as follows: let $\Delta$ denote the Laplace-Beltrami operator on $G/H$. 
One can show that $\Delta\pi^{(\lambda)} = -(\norm{\lambda}^2+\norm{\rho})^2\pi^{(\lambda)}$.
Then, similarly to the Euclidean case, one can apply the spherical Fourier transform to the sides of the heat equation, solve the resulting algebraic equation, and take the inverse spherical Fourier transform to obtain the solution.
\end{proof}

\section{Proofs} \label{apdx:proofs}

\KernelEstimatorVar*

\begin{proof}
Since the estimator is an average of IID random variables, it is enough to estimate the variance of one term.
Write
\[
\Var(\sigma^2 e^{(i\lambda+\rho)^{\top} a(h \bdot g)})
\leq
\E \abs{\sigma^2 e^{(i\lambda+\rho)^{\top} a(h \bdot g)}}^2
=
\sigma^{\mathrlap{2}}
\int_{\fr{a}^*}
\int_H
\abs{e^{(i\lambda+\rho)^{\top} a(h \bdot g)}}^2
\d \mu_H(h) \d \mu_k(\lambda).
\]
Denote the identity element of the group $G$ by $e$ and apply~\Cref{thm:spherical_symmetry} with $g_1 = g_2 = g$ to get
\[
\pi^{(\lambda)}(e)
=
\int_H
e^{(i\lambda+\rho)^{\top} a(h\bdot g)}
\overline{e^{(i\lambda+\rho)^{\top} a(h\bdot g)}}
\d \mu_H(h)
=
\int_H
\abs{e^{(i\lambda+\rho)^{\top} a(h\bdot g)}}^2
\d \mu_H(h).
\]
On the other hand, by definition, we have
\[
\pi^{(\lambda)}(e)
=
\int_H
e^{(i\lambda+\rho)^{\top} a(h)}
\d \mu_H(h).
\]
Since $h = e \bdot e \bdot h$, by the uniqueness of Iwasawa decomposition, we have $A(h) = e$ and therefore $a(h) = 0 \in \fr{a}$, implying $\pi^{(\lambda)}(e) = 1$.
Hence
\[
\Var(\sigma^2 e^{(i\lambda+\rho)^{\top} a(h \bdot g)})
\leq
\sigma^2
\int_{\fr{a}^*}
1
\d \mu_H(h) \d \mu_k(\lambda)
=
\sigma^4.
\]
Since the variance of an IID sum is equal to the sum of variances of its components, the claim follows.
\end{proof}

\RFFConvergence*
\begin{proof}
Recall that the estimator under consideration is
\[
k(g_1 \bdot H, g_2 \bdot H)
&\approx
\frac{\sigma^2}{L}
\sum_{l=1}^L
e^{(i\lambda_l+\rho)^{\top} a(h_l\bdot g_1)}
\overline{e^{(i\lambda_l+\rho)^{\top} a(h_l\bdot g_2)}}
&
\lambda_l &\~ \frac{1}{\sigma^2} \mu_k
&
h_l &\~ \mu_H.
\]
Since it is an average of IID variables, it is enough to estimate variance of only one term.
To this end, write
\[
\label{inq:rff_var}
\Var
\Big(
\sigma^2
e^{(i\lambda+\rho)^{\top} a(h \bdot g_1)}
&\overline{e^{(i\lambda+\rho)a(h \bdot g_2)}}
\Big)
\leq
\E
\abs{
\sigma^2
e^{(i\lambda+\rho)^{\top} a(h \bdot g_1)}
\overline{e^{(i\lambda+\rho)^{\top} a(h \bdot g_2)}}
}^2
\\
&=
\sigma^2
\int_{\fr{a}^*}
\int_H
\abs[0]{
e^{(i\lambda+\rho)^{\top} a(h \bdot g_1)}
e^{(-i\lambda+\rho)^{\top} a(h \bdot g_2)}
}^2
\d\mu_H(h) \d\mu_k(\lambda)
\\
&=
\sigma^2
\int_{\fr{a}^*}
\int_H
e^{2\rho^{\top} a(h \bdot g_1)}
e^{2\rho^{\top} a(h \bdot g_2)}
\d\mu_H(h) \d\mu_k(\lambda)
\\
&=
\sigma^4
\int_H
e^{2\rho^{\top} a(h \bdot g_1)}
e^{2\rho^{\top} a(h \bdot g_2)}
\d\mu_H(h)
\\ \label{eqn:var_sampling_proof_holder}
&\leq
\sigma^4
\del{
\int_H
e^{4\rho^{\top} a(h \bdot g_1)}
\d\mu_H(h)
}^{\frac{1}{2}}
\del{
\int_H
e^{4\rho^{\top} a(h \bdot g_2)}
\d\mu_H(h)
}^{\frac{1}{2}}
.
\]
Recall that for $\lambda \in \fr{a}^* = \R^n$ we have
\[ \label{eqn:apdx:spherical_functions}
\pi^{(\lambda)}(g)
=
\int_H e^{(i\lambda+\rho)^{\top} a(h \bdot g)} \d \mu_H(h).
\]
Let us formally extend the definition of $\pi^{(\lambda)}$ to $\lambda \in \C^n$.
Then~\Cref{eqn:var_sampling_proof_holder} directly implies
\[
\Var
\del{
\sigma^2
e^{(i\lambda+\rho)^{\top} a(h \bdot g_1)}
\overline{e^{(i\lambda+\rho)^{\top} a(h \bdot g_2)}}
}
\leq
\sigma^4
\del{
\pi^{(-3 i \rho)}(g_1)
}^{1/2}
\del{
\pi^{(-3 i \rho)}(g_2)
}^{1/2}.
\]
By \textcite[Chapter IV, \S 2.1]{ggahelgason2000}, the functions $\pi^{(\lambda)}$ are well-defined and continuous for all $\lambda \in \C^n$: more precisely, these functions are \emph{defined} as continuous solutions of certain functional equation, after which \textcite[Ch. IV, \S 4.2, Th. 4.3]{ggahelgason2000} proves that for $\lambda\in\C^n$ they have the form~\Cref{eqn:apdx:spherical_functions}.
Therefore, on any compact $U \subseteq G$ the function $\pi^{(- 3 i \rho)}$ is bounded.
Denoting ${C_U = \sigma^4 \max_{g\in U} \pi^{-3i\.\rho}(g)}$, we get
\[
\Var
\del{
\sigma^2
e^{(i\lambda+\rho)^{\top} a(h \bdot g_1)}
\overline{e^{(i\lambda+\rho)^{\top} a(h \bdot g_2)}}
}
&\leq
\sigma^4
C_U
&
g_1, g_2 &\in U \subseteq G.
\]
This proves the first part of the claim.

To prove the second part of the claim we show that the estimator of the variance, i.e. in the case of $g_1 = g_2 = g \in G$, is globally unbounded.
By the proof of~\Cref{thm:kernel_estimator_var}, we have 
\[
1 = \pi^{\lambda}(e) = \int_H e^{2 \rho^{\top} a(h \bdot g)} \d \mu_H(h) = \E e^{2 \rho^{\top} a(h \bdot g)}
.
\]
Thus
\[
\Var
\del{
\sigma^2
e^{(i\lambda+\rho)^{\top} a(h \bdot g)}
\overline{e^{(i\lambda+\rho)^{\top} a(h \bdot g)}}
}
&=
\Var
\del{
\sigma^2
e^{ 2 \rho^{\top} a(h \bdot g)}
}
\\
&=
\E
\abs{
\sigma^2
e^{ 2 \rho^{\top} a(h \bdot g)}
}^2
-
\abs{
\E
\sigma^2
e^{ 2 \rho^{\top} a(h \bdot g)}
}^2
\\ \label{eqn:var_sampling_proof_unbounded}
&=
\sigma^4
\del{
\pi^{(-3 i \rho)}(g)
-
1
}.
\]
By \textcite[Chapter IV, Theorem 8.1]{ggahelgason2000}, the necessary and sufficient condition for $\pi^{(\lambda)}: G \-> \C$ to be bounded, in the sense $\sup_{g \in G} \abs{\pi^{(\lambda)}(g)} < \infty$, is
\[
\lambda \in \fr{a}^* + i \operatorname{Conv}(\{w\rho : w\in W\})
\]
where $W$ is the Weyl group equipped with its natural action on $\fr{a}^*$---see~\Cref{apdx:noncompact} for the respective definition---and $\operatorname{Conv}$ denotes the convex hull of a set.
Observe that $- 3 i \rho \not\in \fr{a}^* + i \operatorname{Conv}(\{w\rho : w\in W\})$ because $\abs{w \rho} = \abs{\rho}$ and thus $3 \rho \not\in \operatorname{Conv}(\{w\rho : w\in W\})$.
Hence, the right-hand side of~\Cref{eqn:var_sampling_proof_unbounded} is unbounded, which completes the proof.
\end{proof}

\SmoothnessNoncomp*
\begin{proof}
The characterization of Sobolev spaces on a non-compact symmetric space given by \textcite[Section~6]{strichartz1983}---see also \textcite{grosse2013} to clarify equivalence with the classical definition---implies that $f \in H^{\nu+n/2}$ if and only if
\[
\int_{\fr{a}^*} \del{\norm{\lambda}^2 + \norm{\rho}^2}^{\nu+n/2} \abs[0]{\h{f}(\lambda)}^2 \abs{c(\lambda)}^{-2}d\lambda < \infty
\]
where $\h{f}$ is as in~\Cref{eqn:spherical_fourier_transform}.
Denote $f(g) = k_{\nu, \kappa, \sigma^2}(g \bdot H, e \bdot H)$ and $h(g) = \h{k}_{\nu, \kappa, \sigma^2}(g \bdot H, e \bdot H)$.
If we assume $\nu < \infty$, then
\[
\h{f}(\lambda)
\propto
\del{\frac{2 \nu}{\kappa^2} + \norm{\lambda}^2 + \norm{\rho}^2}^{-\nu-n/2}
&&
\h{h}(\lambda)
\propto
\del{\frac{2 \nu}{\kappa^2} + \norm{\lambda}^2}^{-\nu-n/2}.
\]
Direct computation shows
\[
\del{\norm{\lambda}^2 + \norm{\rho}^2}^{-\nu-n/2} c(\nu, \kappa, \rho, n) \leq \abs[0]{\h{f}(\lambda)} \leq C(\nu, \kappa, \rho, n) \del{\norm{\lambda}^2 + \norm{\rho}^2}^{-\nu-n/2}
\\
\del{\norm{\lambda}^2 + \norm{\rho}^2}^{-\nu-n/2} \widetilde{c}(\nu, \kappa, \rho, n) \leq \abs[0]{\h{h}(\lambda)} \leq \widetilde{C}(\nu, \kappa, \rho, n) \del{\norm{\lambda}^2 + \norm{\rho}^2}^{-\nu-n/2}.
\]
Observing that, for any $\nu > 0$, we have
\[
\int_{\fr{a}^*} \del{\norm{\lambda}^2 + \norm{\rho}^2}^{-\nu-n/2} \abs{c(\lambda)}^{-2}d\lambda < \infty
\]
which is shown in the proof of~\Cref{thm:prop_spectral_measure_non_dirac}, implies that $f, h \in H^{\nu+n/2}$.
The functions $k_{\nu, \kappa, \sigma^2}(\., g_2 \bdot H)$, $\h{k}_{\nu, \kappa, \sigma^2}(\., g_2 \bdot H)$ for $g_2 \not=e$ are simple shifts of the functions $f$ and $h$, thus the same property holds for them as well.
Finally, in the case of $\nu=\infty$, the Fourier transform is upper bounded by $C(\alpha, \kappa, \rho, n) \del{\norm{\lambda}^2 + \norm{\rho}^2}^{-\alpha}$ for all $\alpha > 0$, which proves that the respective kernels lie in the Sobolev spaces $H^{\alpha}$ for all $\alpha > 0$.

Since symmetric spaces, as well as all homogeneous spaces with bi-invariant Riemannian metric, are spaces of bounded geometry \cite{eldering2012}, we can use \textcite[Theorem 3.4]{hebey2000} to get the claimed continuous differentiability properties from the appropriate analogs of the classical Sobolev embedding theorems.
\end{proof}

\SnIntegrationLemma*
\begin{proof}
We use the Iwasawa decomposition of \Cref{example:so1n_iwasawa}. 
With this, every $\m{M} \in \SO_0(1,n)$ can be decomposed into the product
\[
\m{M} = \m{N}_{\v{b}} \m{A}_t \m{H} =
\begin{pmatrix}
1 + \Delta & \Delta & \v{b}^\top \\
-\Delta & 1 - \Delta & -\v{b}^\top \\
\v{b} & \v{b} & \m{I}_{n-1}
\end{pmatrix}
\begin{pmatrix}
\cosh(t) & \sinh(t) \\
\sinh(t) & \cosh(t) \\
& & \m{I}_{n-1}
\end{pmatrix}
\m{H}
\]
where $\v{b}\in\mathbb{R}^{n-1}$, $\Delta=\frac{\norm{\v{b}}^2}{2}$, $t\in\mathbb{R}$, $\m{H} \in H$.

Observe that
\[
\begin{pmatrix}
1 \\
& 1 \\
& & \tl{\m{U}}
\end{pmatrix}
\begin{pmatrix}
1 + \Delta & \Delta & \v{b}^\top \\
-\Delta & 1 - \Delta & -\v{b}^\top \\
\v{b} & \v{b} & \m{I}_{n-1}
\end{pmatrix}
& =
\begin{pmatrix}
1 + \Delta & \Delta & \v{b}^\top \\
-\Delta & 1 - \Delta & -\v{b}^\top \\
\tl{\m{U}}\v{b} & \tl{\m{U}}\v{b} & \tl{\m{U}}
\end{pmatrix}
\\ & =
\begin{pmatrix}
1 + \Delta & \Delta & (\tl{\m{U}}\v{b})^\top \\
-\Delta & 1 - \Delta & -(\tl{\m{U}}\v{b})^\top \\
\tl{\m{U}}\v{b} & \tl{\m{U}}\v{b} & \m{I}_{n-1}
\end{pmatrix}
\begin{pmatrix}
1 \\
& 1 \\
& & \tl{\m{U}}
\end{pmatrix}.
\]
Also note that $\m{U}$ commutes with $\m{A}_t$, i.e. $\m{U} \m{A}_t = \m{A}_t \m{U}$.
Using these two properties, we infer that $\m{U}\m{M} = \m{N}_{\tl{\m{U}}\v{b}} \. \m{A}_t \. \m{U}\m{H}$.
Hence, by the uniqueness of the Iwasawa decomposition, we have
\[
N(\m{U}\m{M}) &= \m{N}_{\tl{\m{U}}\v{b}}
&
A(\m{U}\m{M}) &= \m{A}_t
&
H(\m{U}\m{M}) &= \m{U}\m{H}.
\]
In particular, $A(\m{U}\m{M}) = \m{A}_t = A(\m{M})$, which proves the claim.
\end{proof}

\HeatHypSamp*
\begin{proof}
The densities $p_{x_j}$ of $x_j$ and $p_{y_j}$ of $y_j$ are given by
\[
p_{x_j}(\xi)
&=
2^{\frac{1-j}{2}}
\Gamma\del{\frac{j+1}{2}}^{-1}
\xi^j
e^{-\xi^2/2}
&
p_{y_j}(\psi)
&=
p_{x_j}(\kappa \psi)\kappa
\]
where the second expression follows from the change of variables formula~for~densities.
Write
\[
p_{\infty,\kappa}(\lambda)
\propto
\sum_{j=0}^{m}
\alpha_j \lambda^{j}
e^{-\frac{\kappa^2}{2}\lambda^2}
=
\sum_{j=0}^{m}
\alpha_j
\beta_j^{-1}
\ubr{
\beta_j
\lambda^{j}
e^{-\frac{\kappa^2}{2}\lambda^2}
}_{p_{y_j}(\lambda)}
\propto
\sum_{j=0}^{m}
c_j
p_{y_j}(\lambda)
.
\]
The right-hand side is an actual density, which corresponds to the mixture of $p_{y_j}$ with coefficients $c_j$, which proves the claim.
\end{proof}

\MaternHypSamp*
\begin{proof}
We have $y_j = \Phi(x_j)$ with $\Phi: [0, \infty) \-> [0, \infty)$ a bijective map given by $\Phi(\xi) = \sqrt{\gamma \xi}$.
Then, using the change of variable formula for densities, write
\[
p_{y_j}(\psi)
&=
p_{x_j}(\psi^2/\gamma) 2 \psi / \gamma
&
p_{x_j}(\xi)
&=
\frac{\xi^{\frac{j-1}{2}}(1+\xi)^{-\nu-\frac{n}{2}}}{B(\frac{j+1}{2},\nu+\frac{n-j-1}{2})}.
\]
Then
\[
p_{y_j}(\psi)
&=
\frac{\del{\psi^2/\gamma}^{\frac{j-1}{2}} (1+\psi^2/\gamma)^{-\nu-\frac{n}{2}}}{B\del[2]{\frac{j+1}{2}, \nu+\frac{n-j-1}{2}}}
2 \psi / \gamma
=
\frac{2 \gamma^{\nu+\frac{n-j-1}{2}} \psi^{j}(\gamma+\psi^2)^{-\nu-\frac{n}{2}}}{B\del[2]{\frac{j+1}{2},\nu+\frac{n-j-1}{2}}}
\\
&=
\ubr{
2
B\del[2]{\frac{j+1}{2},\nu+\frac{n-j-1}{2}}^{-1}
\gamma^{\nu+\frac{n-j-1}{2}}
}_{\beta_j}
\psi^{j}
\del{\gamma+\psi^2}^{-\nu-\frac{n}{2}}
.
\]
It follows that
\[
p_{\nu,\kappa}(\lambda)
\propto
\sum_{j=0}^m
a_j
\lambda^{j}
\del{
\gamma
+
\lambda^2
}^{-\nu-n/2}
=
\sum_{j=0}^m
a_j
\beta_j^{-1}
\ubr{\beta_j
\lambda^{j}
\del{
\gamma
+
\lambda^2
}^{-\nu-n/2}}_{p_{y_j}(\lambda)}
\propto
\sum_{j=0}^m
c_j
p_{y_j}(\lambda)
.
\]
The right-hand side is an actual density, which corresponds to the mixture of $p_{y_j}$ with coefficients $c_j$, which proves the claim.
\end{proof}

\SPDHomogeneous*
\begin{proof}
Recall that every symmetric positive-definite matrix $\m{S}\in \SPD(d)$ admits a Cholesky decomposition $\m{S}=\m{C}\m{C}^\top$ where $\m{C}$ is a lower-triangular matrix.
The map 
\[
\GL(d) &\-> \SPD(d)
&
\m{M} &\|> \m{M} \lacts \m{I}_{d} = \m{M}\m{M}^{\top}
\]
is thus surjective. 
For $\m{S}_1$ and $\m{S}_2\in \SPD(d)$ with Cholesky factors $\m{C}_1$ and $\m{C}_2$, respectively, 
\[
\label{eq:psd:G_translation}
\m{C}_2\m{C}_1^{-1} \lacts \m{S}_1
=
\m{C}_2 \m{C}_2^{\top}
=
\m{S}_2
\]
hence this group action is transitive.
Note in particular that 
\[
\m{H} \lacts \m{I}_d =\m{H}\m{H}^{\top} = \m{I}_d
\]
holds if and only if $\m{H}\in \Ort(d)$.
Thus the isotropy group of the element $\m{I}_d$ is exactly $\Ort(d)$.
This means that $\SPD(d)$ may be regarded as the quotient $\GL(\R^d) / \Ort(d)$, which is a smooth homogeneous space.
The identity $s(s(\m{M})) = s$ is immediate.
The condition $s(\m{M}) = \m{M}$ means $\m{M^{-\top}} = \m{M}$ and is equivalent to $\m{M} \m{M}^\top = \m{I}_d$: hence it is true if and only if $\m{M} \in \Ort(d)$.
\end{proof}

\SSPDSymmetric*
\begin{proof}
First, we restrict the action $\lacts$ defined by~\Cref{eqn:spd_action} and the involution $s$ defined by~\Cref{eqn:spd_s} from the group $\GL(d)$ to its subgroup $\SL(d)$ consisting of unit-determinant matrices.
Applying the proof of \Cref{thm:spd_homogeneous} \emph{mutatis mutandis} allows us to regard $\SSPD(d)$ as the smooth homogeneous space $\SL(d)/\SO(d)$.
Using the fact that $\SL(d)$, in contrast to $\GL(d)$, is connected, we conclude that $\SSPD(d)$ is a symmetric space.
We now verify \Cref{asm:symmetric_space_assumptions}.
\1 By standard theory, the group $\SL(d)$ for $d > 1$ is a simple Lie group, and hence semi-simple \cite[Ch.~III, \S8]{helgason1979differential}.
\2 The center of $\SL(d)$ consists of the scalar matrices corresponding to $d$th roots of unity in the base field \cite[Theorem~8.9]{rotman1995}, which is a finite set.
\3 $\SO(d)$ is a maximal compact subgroup of $\SL(d)$: every compact subgroup $K$ of $\GL(d)$ preserves some inner product---to find one, take an arbitrary inner product and average it over $K$---while the orthogonal groups of all inner products in $d$-dimensional space are isomorphic, and a Lie group cannot have a proper Lie subgroup with the same dimension and number of connected components.
\0
The claim follows.
\end{proof}

\SPDAsProduct*

\begin{proof}
From the definition of $\c{T}$, it is clear that this is a diffeomorphism.
We will show that this map preserves geodesic distance, which by virtue of the Myers--Steenrod theorem \cite{myers1939} implies that $\c{T}$ is a Riemannian isometry.

Recall that the canonical metric on $\SSPD(d)$ is defined by pushforward of the bilinear form on the tangent space at $\m{I}_d$ that itself is induced by the Killing form.
The geodesic distance induced by this metric is known to correspond to the geodesic distance $d_{\SSPD(d)}$ coinciding with the right-hand side of~\Cref{eqn:affine_inv_dist}. Indeed, the identification of the metric on $\SPD(d)$ \cite[Lemma~6.1.1]{bhatia2009} allows one to show that the unique geodesic between two unit-determinant matrices lies inside $\SSPD(d)$ \cite[Theorem~6.1.6]{bhatia2009}.

Since the Frobenius norm of a symmetric matrix is the $l^2$-norm of the vector of its eigenvalues, both $d_{\SPD(d)}$ and $d_{\SSPD(d)}$, for a pair of appropriate matrices $\m{S}_1$ and $\m{S}_2$, are equal to
\[
\del{\sum_{j=1}^d \log^2 \lambda_j}^{1/2}
\]
where $\lambda_1,\lambda_2,\ldots,\lambda_d$ are eigenvalues of $\m{S}_1^{-1/2}\m{S}_2\m{S}_1^{-1/2}$.

The canonical metric on $\R_{>0}$ corresponds to the distance $d_{\R_{>0}}(a_1,a_2) = \abs{\log(v_1 v_2^{-1})} =  \abs{\log(v_1)-\log(v_2)}$, which is the pullback of the Euclidean distance through the logarithm.

From this, the remaining argument is straightforward. 
Take $\m{S}_1, \m{S}_2\in \SSPD(d)$ and $a_1, a_2 \in \R_{>0}$.
Denote the eigenvalues of $\m{S}_1^{-1/2}\m{S}_2\m{S}_1^{-1/2}$ by $\lambda_1,\lambda_2,\ldots,\lambda_d$.
Note that in this case $\sum_{j=1}^d \log \lambda_j = 0$, since the determinant is equal to $1$, hence 
\[
\f{dist}_{\SPD(d)}(v_1\m{S}_1,v_2\m{S}_2)^2
&=
\|v_1v_2^{-1} \m{S}_1^{-1/2} \m{S}_2 \m{S}_1^{-1/2}\|_F^2 
\\
&= 
\sum_{j=1}^d \log(v_1v_2^{-1}\lambda_j)^2
\\
&=
\sum_{j=1}^d \log(\lambda_j)^2 + 2\log\del{\frac{v_1}{v_2}}\sum_{j=1}^d \log\lambda_j + \log\del{v_1^{\sqrt{d}} v_2^{-\sqrt{d}}}^2 
\\
&=
d_{\SSPD(d)}(\m{S}_1,\m{S}_2)^2+d_{\R_{>0}}(v_1^{\sqrt{d}},v_2^{\sqrt{d}})^2.
\]
The last equation is exactly the square of the product metric between the pairs $(a_1, \m{S}_1)$ and $(a_2, \m{S}_2)$, which completes the proof.
\end{proof}

\begin{CustomLemma}
\label{lem:chi-distribution-calculation}
Consider a random vector $\v{x}$ on $\R^d$ with density $p_{\v{x}}(\v\xi) \propto e^{- \gamma \norm{\v\xi}^2/2} f(\v\xi)$ where $\gamma > 0$ and $f: \R^d \-> [0, \infty)$ satisfies $f(a \v\xi) = a^k f(\v\xi)$ for some $k \in \N$ and all $a > 0$.
Let $y$ be a random variable independent from~$\v{x}$, distributed according to the~$\chi_{2 \nu}$ distribution, namely $p_{y}(\psi) \propto \psi^{2 \nu-1} e^{-\psi^2/2}$ and $p_{\v{x}, y}(\v\xi, \psi) = p_{\v{x}}(\v\xi)p_{y}(\psi)$.
Then the random vector $\v{z} = \v{x}/y$ has the following density:
\[
p_{\v{z}}(\v\zeta) &\propto f(\v\zeta)(\gamma^{-1}+\norm{\v\zeta}^2)^{-\alpha}
&
\alpha &= \nu + (d+k)/2
.
\]
\end{CustomLemma}

\begin{proof}
Let us compute the joint density $p_{\v{z}, y}$.
We have $(\v{z}, y) = \v{\Phi}(\v{x}, y)$, where the function $\v{\Phi}: \R^d \x \R_{>0} \-> \R^d \x \R_{>0}$ is given by $\v{\Phi}(\v\xi, \psi) = (\v\xi/\psi, \psi)$.
Since $\v{\Phi}$ is bijective, we can use the change of variables formula for densities to write
\[
p_{\v{z}, y}(\v\zeta, \psi)
=
p_{\v{x}, y}(\psi \v\zeta, \psi)
\det\del{\m{J}_{\v{\Phi}^{-1}}(\v\zeta, \psi)}
\]
where $\m{J}_{\v{\Phi}^{-1}}$ is the Jacobian matrix of the map $\v{\Phi}^{-1}: \R^d \x \R_{>0}$ given by $\v{\Phi}^{-1}(\v\zeta, \psi) = (\psi \v\zeta , \psi)$.
It is easy to see that
\[
\m{J}_{\v{\Phi}^{-1}}(\v\zeta, \psi)
&=
\begin{pmatrix}
\psi & 0 & \dots & 0 & \zeta_1 \\
0 & \psi & \dots & 0 & \zeta_2 \\
\vdots &  & \ddots &  & \vdots \\
0 & 0 & \dots & \psi & \zeta_d \\
0 & 0 & \dots & 0 & 1 \\
\end{pmatrix}
&
\det\del{\m{J}_{\v{\Phi}^{-1}}(\v\zeta, \psi)}
&=
\psi^d.
\]
Thus
\[
p_{\v{z}, y}(\v\zeta, \psi)
=
p_{\v{x}, y}(\psi \v\zeta, \psi)
\psi^d
=
p_{\v{x}}(\psi \v\zeta)
p_{y}(\psi)
\psi^d
&\propto
e^{-\gamma\norm{\psi \v\zeta}^2/2}
\psi^k f(\v\zeta)
\psi^{2\nu-1} e^{-\psi^2/2}
\psi^d
\\
&\propto
\psi^{d+k+2\nu-1}
e^{-\psi^2/2\del{\gamma\norm{\v\zeta}^2 + 1}}
f(\v\zeta)
.
\]
Now, we can compute the marginal distribution of $\v{z}$, up to a constant, by writing

\[
p_{\v{z}}(\v\zeta)
&=
\int_{\R_{>0}}
p_{\v{z}, \psi}(\v\zeta, \psi)
\d\psi
\propto
f(\v\zeta)
\int_{\R_{>0}}
e^{-\psi^2\del{\gamma\norm{\v\zeta}^2+1}/2}
\psi^{d+k+2\nu-1}
\d \psi
\\
&\propto f(\v\zeta)(\gamma\norm{\v\zeta}^2+1)^{-(d+k+2\nu)/2}
\propto
f(\v\zeta)(\gamma^{-1} + \norm{\v\zeta}^2)^{-(d+k+2\nu)/2}
\]

where the integral was computed using \textcite[Section 3.326, Item 2]{gradshteyn2014}.
This finishes the proof of the auxiliary statement.
\end{proof}

\SPDSampling*

\begin{proof}
Applying \nameCref{lem:chi-distribution-calculation}, taking $p_{\v{x}}$ equal to $p_{\infty,\gamma}$, where
\[
p_{\infty,\gamma}(\v\lambda)
&\propto
\exp\del{-\del{\frac{2 \nu}{\kappa^2} + \norm{\v{\rho}}^2}^{-1}(\norm{\v{\lambda}}^2+\norm{\v{\rho}}^2)/2}
\prod\limits_{1\le i<j\le d}
\pi |\lambda_i-\lambda_j|
\\
&\propto
\exp\del[3]{
-
\ubr{
\del{\frac{2 \nu}{\kappa^2} + \norm{\v{\rho}}^2}^{-1}
}_{\gamma^2}
\norm{\v{\lambda}}^2/2
}
\ubr{
\prod\limits_{1\le i<j\le d}
\pi |\lambda_i-\lambda_j|
}_{f(\v\lambda)}.
\]
We have $f(a \v\lambda) = a^{\frac{d(d-1)}{2}} f(\v\lambda)$.
Then, \nameCref{lem:chi-distribution-calculation} shows that for a chi-distributed random variable $y \sim \chi_{2\nu}$, the random vector $\v\lambda^{\nu,\kappa} = \v\lambda^{\infty,\gamma}/y$ has density
\[
p_{\v{\lambda}^{\nu,\kappa}}(\v\lambda)
&\propto
f(\v{\lambda})
(\gamma^{-2} + \abs{\v\lambda}^2)^{-(d+\frac{d(d-1)}{2} + 2\nu)/2}
\\
&=
\del{
\frac{2 \nu}{\kappa^2} + \norm{\v\lambda}^2 + \norm{\v{\rho}}^2
}^{-(\nu + \frac{d(d+1)}{4})}
\prod_{1\le i<j\le d}
\pi |z_i-z_j|
.
\]
From this, and noting that the dimension of $\SPD(d)$ is $n=\frac{d(d+1)}{2}$, the claim follows.
\end{proof}
\PropKernelUpperBound*
\begin{proof}
Recall the definition of zonal spherical functions $\pi^{\del{\lambda}}$, namely
\[
\pi^{\del{\lambda}}(g) = \int_H e^{\del{i\lambda + \rho}^{\top} a\del{h \bdot g}}\d\mu_H(h).
\]
Write
\[
\abs[0]{\pi^{\del{\lambda}}(g)}
\leq
\int_H \abs[0]{e^{\del{i\lambda + \rho}^{\top} a\del{h \bdot g}}}\d\mu_H(h)
=
\int_H e^{\rho^{\top} a\del{h \bdot g}}\d\mu_H(h) = \pi^{(0)}(g).
\]
Let $\mu_k$ be the spectral measure of the kernel $k$.
Then
\[
\abs{\Bbbk(g)}
=
\abs{k(g \bdot H, e)}
=
\abs{\int_{\fr{a}^*} \pi^{\del{\lambda}}(g)\d\mu_k(\lambda)}
\leq
\int_{\fr{a}^*} \abs[0]{\pi^{\del{\lambda}}(g)}\d\mu_k(\lambda) = \Bbbk(e)\pi^{\del{0}}(g)
\]
because $\pi^{\del{\lambda}}(e) = 1$, and therefore
\[
\Bbbk(e) = \int_{\fr{a}^*} \pi^{\del{\lambda}}(e)\d\mu_k(\lambda)
=
\int_{\fr{a}^*} 1 \d\mu_k(\lambda)
.
\]
Finally, note that $\pi^{\del{0}} \neq 1$, since $\Delta \pi^{\del{0}} = -\norm{\rho}^2\pi^{\del{0}}$ for $\rho \neq 0$, while $\Delta 1 = 0$.
What is more, $\pi^{\del{0}}(g) \leq \pi^{\del{0}}(e) = 1$ since $g_1, g_2 \mapsto \pi^{\del{0}}(g_2^{-1} \bdot g_1)$ is a positive semi-definite function.
\end{proof}
\PropSpectralMeasureNonDirac*

\begin{proof}
Recall that 
\[
\mu_{\nu, \kappa, \sigma^2}(\lambda)
&\propto
\del{\frac{2 \nu}{\kappa^2} + \norm{\lambda}^2 + \norm{\rho}^2}^{-\nu-n/2} \abs{c(\lambda)}^{-2}d\lambda
\\
\mu_{\infty, \kappa, \sigma^2}(\lambda) &\propto e^{-\frac{\kappa^2}{2}\del{\norm{\lambda}^2 + \norm{\rho}^2}}\abs{c(\lambda)}^{-2}d\lambda
.
\]
The proof is based on analysis of Harish-Chandra's $c$-function. 
This function is meromorphic and has \cite[p. 109]{helgason1994} the following form
\[
\label{eq:formula_on_c_function}
c(\lambda) 
\propto 
\prod_{\alpha\in \Sigma^+}
\frac{
    2^{-i \lambda^{\top} \alpha/\norm{\alpha}^2}
    \Gamma\del{i \lambda^{\top} \alpha/\norm{\alpha}^2}
}
{
\Gamma\del{
    \frac{1}{2}
    \del{\frac{1}{2}m_\alpha+1+i \lambda^{\top}\alpha/\norm{\alpha}^2}
    }
\Gamma\del{
    \frac{1}{2}
    \del{\frac{1}{2}m_\alpha+m_{2\alpha}+i \lambda^{\top} \alpha/\norm{\alpha}^2}
    }
}
\]
where the product is taken over the the set of positive roots $\Sigma^+$ and $m_\alpha$ denotes the multiplicity of the root $\alpha$, see \Cref{apdx:noncompact} for a brief introduction into these notions.\footnote{Our notion of positive roots coincides with the notion of \emph{indivisible} positive roots in \textcite{helgason1994}.}

By \textcite[Lemma 6.2]{strichartz1983}, there exists a constant $c_0 > 0$ such that for all $s > 1$
\[
    \label{eq:bounds_on_c_function}
    \int_{\norm{\lambda} \le s} \abs{c(\lambda)}^{-2}d\lambda \le c_0s^n.
\]
With these expressions, we are ready to prove the proposition.
Without loss of generality, we assume $\sigma^2 = 1$.
We prove the first part of statement by showing that, as $\kappa \rightarrow \infty$, the measures $\mu_{\nu, \kappa, 1}$ converge to the measure
\[
\mu_{\nu,\infty}(\lambda)
&\propto
\del{\norm{\lambda}^2 + \norm{\rho}^2}^{-\nu-n/2} \abs{c(\lambda)}^{-2}\d\lambda
\]
rather than to a Dirac delta function.
We start by showing that $\mu_{\nu,\infty}$ is a finite measure.

Applying \Cref{eq:bounds_on_c_function}, we bound
\[
\int_{s \leq \norm{\lambda} \leq 2s}
1 \d\mu_{\nu, \infty}
&\propto
\int_{s \leq \norm{\lambda} \leq 2s}
\del{\norm{\lambda}^2 + \norm{\rho}^2}^{-\nu-n/2}
\abs{c(\lambda)}^{-2}\d\lambda
\\
&\leq
\del{s^2 + \norm{\rho}^2}^{-\nu-n/2}
\int_{\norm{\lambda} \leq 2s}
\abs{c(\lambda)}^{-2}\d\lambda
\lesssim
s^{-n-2\nu} s^n
=
s^{-2\nu}
\]
where $\lesssim$ denotes inequality up to a constant.
Using this, we can write
\[
\int_{\fr{a}^*}
1 \d\mu_{\nu, \infty}
= 
\int_{\norm{\lambda} \le 1}
1 \d\mu_{\nu, \infty}
+
\sum_{k=0}^\infty
\int_{\mathrlap{2^k \leq \norm{\lambda} < 2^{k+1}}}
\qquad
1 \d\mu_{\nu, \infty}
\lesssim 
c_0 \norm{\rho}^{-2\nu-n} + \sum_{k=0}^{\infty} 2^{-2\nu k}
<
\infty.
\]
This proves that the measure $\mu_{\nu,\infty}$ is finite.

When $\kappa \rightarrow \infty$, the densities of $\mu_{\nu, \kappa, 1}$ with respect to the Lebesgue measure $\d\lambda$ converge to the density of $\mu_{\nu, \infty}$ pointwise.
Thus, by Scheffé's lemma, recalling that~$\d\mu_{\nu, \infty}$ is finite, the measures $\mu_{\nu, \kappa, 1}$ converge in distribution to $\mu_{\nu, \infty}$.
This proves the first part of the statement.

Now, we prove the second part of the statement.
For this, we will show that if either $\nu = \infty$, or if we instead work with $\h{\mu}_{\nu,\kappa, \sigma^2}$ and $\nu$ is large enough, the mass of the measures concentrates at $\lambda  = 0$ as $\kappa \rightarrow \infty$.
We prove this only for $\h{\mu}_{\nu,\kappa, \sigma^2}$, $\nu < \infty$, as the case of $\nu = \infty$ is similar, but simpler.
As before, we also assume $\sigma^2 = 1$ without any loss of generality.
First, write
\[
\int_{\norm{\lambda} \geq 1} 
\del{\frac{2\nu}{\kappa^2} + \norm{\lambda}^2}^{-\nu-n/2}
\abs{c(\lambda)}^{-2}\d\lambda 
&\leq
\int_{\norm{\lambda} \geq 1} 
\norm{\lambda}^{-2\nu-n}
\abs{c(\lambda)}^{-2}\d\lambda
\\
&=
\sum_{k=0}^{\infty}
\int_{\mathrlap{2^k \leq \norm{\lambda} < 2^{k+1}}}
\qquad
\norm{\lambda}^{-2\nu-n}
\abs{c(\lambda)}^{-2}\d\lambda
\\
&\leq
\sum_{k=0}^{\infty}
2^{-2 \nu k - n k}
\,\,
2^{(k+1)n}
\leq
c_1
<
\infty
.
\]
Next, because of~\Cref{eq:bounds_on_c_function}, we have for any $\varepsilon>0$ that
\[
\int_{\norm{\lambda} \geq \varepsilon} 
\del{\frac{2\nu}{\kappa^2} + \norm{\lambda}^2}^{-\nu-n/2}
\abs{c(\lambda)}^{-2}d\lambda
&\leq
c_1 + \int_{1 > \norm{\lambda} \geq \varepsilon} 
\varepsilon^{-2 \nu - n}
\abs{c(\lambda)}^{-2}
d\lambda
\\
&\leq
c_1 + c_0 \varepsilon^{-2 \nu - n}
<
\infty.
\]
Let us focus on the case $\norm{\lambda} < \varepsilon$.
The formula for $c(\lambda)$ gives us that, in this case, ${\abs{c(\lambda)}^{-2} \approx c_3 \prod_{\alpha \in \Sigma^+} \abs{\lambda^{\top}\alpha}^{2}}$, since $\abs{\Gamma(x i)}^{-2} = \frac{x\sinh(\pi x)}{\pi} = x^2 + \c{O}(x^4)$.
So, we have
\[
\norm{\lambda}^{-2\nu-n}\abs{c(\lambda)}^{-2} \approx c_3\norm{\lambda}^{-2\nu-n} \prod_{\alpha \in \Sigma^+}\abs[1]{\lambda^{\top} \alpha}^{2}
\]
and therefore, if $-2 \nu - n + 2 \abs{\Sigma^+} + \dim \fr{a}^* < 0$, the integral 
\[
\int_{\norm{\lambda} < \varepsilon}
\norm{\lambda}^{-2\nu-n}
\prod_\alpha\abs{\lambda^{\top} \alpha}^{2}
\d\lambda
\]
does not converge.
This condition must hold for large enough $\nu$.
We can even be more specific and give an explicit sufficient condition for this: note that since $m_{\alpha} \in \Z_{> 0}$, we have $\sum_{\alpha \in \Sigma^+} m_{\alpha} \geq \sum_{\alpha \in \Sigma^+} 1 = \abs{\Sigma^+}$. Also, by \textcite[p. 409]{strichartz1983}, we have $n = \sum m_{\alpha} + \dim \fr{a}^*$. Hence $-2 \nu - n + 2 \abs{\Sigma^+} + \dim \fr{a}^* \leq -2 \nu - n + \abs{\Sigma^+} + \sum_{\alpha \in \Sigma^+} m_{\alpha} + \dim \fr{a}^* = -2 \nu + \abs{\Sigma^+}$. Thus $\nu > \abs{\Sigma^+}/2$ is a sufficient---but not always necessary---condition for the integral to diverge.
Because of the above, for any fixed $\varepsilon > 0$, as $\kappa \-> \infty$, we have
\[
&\int_{\abs{\lambda} \geq \varepsilon} 
\del{\frac{2\nu}{\kappa^2} + \norm{\lambda}^2}^{\mathrlap{-\nu-n/2}}
\quad
\abs{c(\lambda)}^{-2}
\d\lambda
\leq C_{\varepsilon}
&
&\int_{\abs{\lambda} < \varepsilon} 
\del{\frac{2\nu}{\kappa^2} + \norm{\lambda}^2}^{\mathrlap{-\nu-n/2}}
\quad
\abs{c(\lambda)}^{-2}
\d\lambda
\to
\infty
.
\]
Since $\h{\mu}_{\nu, \kappa, 1}$ is a probability measure, namely $\h{\mu}_{\nu, \kappa, 1}(\fr{a}^*) = 1$, we have
\[
\h{\mu}_{\nu, \kappa, 1}(\cbr{\lambda \in \fr{a}^*: \norm{\lambda} \geq \varepsilon}) \-> 0
&&
\h{\mu}_{\nu, \kappa, 1}(\cbr{\lambda \in \fr{a}^*: \norm{\lambda} < \varepsilon}) \-> 1
\]
which proves that $\h{\mu}_{\nu, \kappa, 1} \-> \delta$ as $\kappa \-> \infty$.
\end{proof}

\printbibliography

\end{document}